\documentclass[a4paper,fleqn,usenatbib]{mnras}
\usepackage{newtxtext,newtxmath}

\usepackage[T1]{fontenc}
\usepackage{ae,aecompl}

\usepackage{graphicx}	
\usepackage{amsmath}	
\usepackage{longtable}
\usepackage{ltxtable}
\usepackage{lscape}
\usepackage{rotating}
\usepackage{multicol}
\usepackage{enumitem}
\usepackage{color}
\usepackage{cite}
\usepackage{natbib} 
\usepackage{journal_names}
\usepackage{afterpage}
\usepackage{caption}
\usepackage[flushleft]{threeparttable}
\usepackage{subfig}
\usepackage{multirow}

\title[Variability in massive protostars]{A Multi-Epoch Study of Radio Continuum Emission from Massive Protostars}

\author[W. O. Obonyo et al.]{
W. O. Obonyo,$^{1}$\thanks{E-mail: pywoo@leeds.ac.uk}
S. L. Lumsden,$^{1}$
M. G. Hoare,$^{1}$
S.E. Kurtz, $^{2}$ 
and S.J.D Purser $^{3}$
\\
$^{1}$School of Physics and Astronomy,
The University of Leeds,
Woodhouse Lane, Leeds LS2 9JT, United Kingdom.
\\
$^{2}$Instituto de Radioastronom\'ia y Astrof\'isica, Universidad Nacional Aut\'onoma de M\'exico, Apartado Postal.\\
$^{3}$Astronomy \& Astrophysics, Dublin Institute for Advanced Studies (DIAS), 10 Burlington Road, Dublin 4, Ireland.
}

\date{Accepted XXX. Received YYY; in original form ZZZ}

\pubyear{2015}

\begin{document}
\label{firstpage}
\pagerange{\pageref{firstpage}--\pageref{lastpage}}

\maketitle

\begin{abstract}
We report the results of the Jansky Very Large Array (JVLA) observation of five massive protostars at 6\, and 22.2\,GHz.
The aim of the study was to compare their current fluxes and positions with previous observations to search for evidence of variability.
Most of the observed sources present the morphologies of a thermal core, hosting the protostar and exhibiting no proper motion, and associated non-thermal radio lobes that are characterised by proper motions and located away from the thermal core. Some of the protostars drive jets whose lobes have dissimilar displacement vectors, implying precession of the jets or the presence of multiple jet drivers. 
The jets of the protostars were found to have proper motions that lie in the range 170$ \leq v \leq $650\,kms$^{-1}$, and precessions of periods 40$\leq p \leq$50\,years and angles 2$\leq \alpha \leq$10$^\circ$, assuming that their velocities $v=$500\,kms$^{-1}$. The core of one of the sources, S255 NIRS 3, which was in outburst at the time of our observations, showed a significant change in flux compared to the other sources. Its spectral index decreased during the outburst, consistent with the model of an expanding gas bubble. Modelling the emission of the outburst as that of a new non-thermal lobe that is emerging from a thermal core whose emission enshrouds that of the lobe also has the potential to account for the increase in flux and a decrease in the spectral index of the source's outburst.\\ 

\end{abstract}

\section{Introduction}
Multi-epoch study of astrophysical objects is crucial in identifying any changes that they display. Massive protostellar jets, for example W75N\,B(\citealt{2015Sci...348..114C}), show evidence of changes in both size and shape. Indeed, protostars are known to manifest variability in a number of ways (\citealt{1995ApJ...449..184M}). They show displacement through tidal interactions with their binary companions \citep{2016A&A...593A.132P} and jet proper motions. Knowledge of variability of the objects can thus be instrumental in; identifying their driving sources and associated knots \citep{2009A&A...502..579C}, estimating the rates at which the jets accelerate and/or decelerate, and providing clues to driving mechanisms and close binary formations.

Unlike in low-mass (\citealt{1996AJ....111..320S}, \citealt{2006A&A...446..155F}) and very low-luminosity protostars \citep{2014ApJ...789....9C} where flux variability has been observed in $\sim$20\%-30\% of the sources, only a few massive protostars are known to manifest variability, some of which may be associated with jet properties. An infrared study of variability in protostars by \citet{2017MNRAS.465.3011C} showed that the changes may be linked to variation in accretion rates. Indeed, a number of studies attribute flux variability to periods of accretion burst (\citealt{2009ApJ...694.1045Z}, \citealt{2006ApJ...650..956V}) that occur within relatively stable rates of inflows. Such accretion activities should manifest through their jets as outflows are tightly linked to inflows. A likely evidence for accretion burst in massive protostars is the discrepancy between their observed and theoretical estimates of accretion rates (\citealt{2016MNRAS.460.1039P}, \citealt{2009Sci...323..754K}). Intervals of high accretion rates, in the form of bursts, could be a key aspect of stellar formation.

Massive protostars can experience rare but intense outbursts of radio emission. The phenomenon was reported in two 20$\,M_\odot$  MYSOs, S255 NIRS 3 \citep{2017NatPh..13..276C} and V723 Carinae \citep{2015MNRAS.446.4088T}. Other objects known to be variable at different wavelengths include NGC 6334I \citep{2017ApJ...837L..29H}, V645 Cygni \citep{2006A&A...457..183C} and G323.46-0.08 \citep{2019MNRAS.487.2407P}.
Such outbursts are relatively rare in massive stars and there is considerable scope for a large scale survey. However, the number of objects of interest vis-a-vis the available instruments does not allow for systematic monitoring over long periods of time, only permitting sporadic multi-epoch observations (\citealt{2018A&A...619A..41T}, \citealt{2013AAS...22125626T}, \citealt{1995ApJ...449..184M}).
Given the value of multi-epoch observations, a sample of five massive protostars, previously observed in 2012 was re-observed in March 2018 to search for and study variability in jets of massive protostars.

\section{Observation and Data}
\subsection{Source selection}
Four objects were selected from a 2012 C-band survey (\color{blue}Purser et al. 2020 in prep\color{black}) for re-observation at the same frequency band in 2018 to search for evidence of positional and/flux variability. The identified sources are spatially resolved at C-band and have lobes with measurable proper motions. Besides the four objects; G133.7150+01.2155, G173.4839+02.4317, G192.6005-00.0479 and G196.4542-01.6777, hereafter G133.7150, G173.4839, G192.6005 and G196.4542 respectively, a fifth source, G173.4815+02.4459, hereafter G173.4815, was observed within the primary beam of G173.4839. 

The time difference between the two observations, approximately five and a half years, allowed for detection of displacements within the jets at the sources' distances of 2$-$5\,kpc. For example, a jet located within a distance d$=$2.0$-$5.0\,kpc, driving out mass at a typical velocity v$=$500\,km\,s$^{-1}$ (\citealt{1995ApJ...449..184M}; \citealt{2006ApJ...638..878C}) should manifest an angular shift of $\theta_s \sim $0.12$-$0.3$^{\prime\prime}$ between the two epochs. Assuming an astrometric precision $\sim \frac{FWHM}{SNR}$ (\citealt{1997ApJ...488..675W}, \citealt{2010arXiv1001.0305M}), a 1$\sigma$ shift of 0.03$^{\prime\prime}$ should be detectable for components with a SNR=10 and a $0.3^{\prime\prime}$ beam. The difference in the epochs also provides an opportunity to search for evidence for flux variability and precession, features that were noted by \color{blue}Purser et al. \color{black}(\color{blue}2020 in prep\color{black}). The thermal cores were also selected for observation at 22.2\,GHz to allow for a more accurate estimate of their free-free SEDs, free from effects of dust and non-thermal emission. Again, the estimated spectral indices would not be subject to variability-related errors.
Table \ref{Table_variability_sample} shows a list of the sources and some of their properties. 
\begin{table*}
    \centering
        \caption{A table of RMS\textsuperscript{*} names, positions, bolometric luminosities and distances of the objects. Last column shows the references for the adopted distances. The references of the distances are shown in the footnote.}
\resizebox{\linewidth}{!}{%
    \begin{tabular}{l l l l l l l}
       \hline
         Source& Common & RA(J2000) &DEC(J2000) & L$_{bol}$ & \multicolumn{2}{c}{Distance}  \\
          & Name & & & (L$_{\odot}$) & (kpc)\\
        \hline
        $\mathrm{G133.7150}$ & W3 IRS5\textsuperscript{e} &02$^h$25$^{\prime}$40.77$^{\prime\prime}$ & +62$^{\circ}$05$^{\prime}$52.4$^{\prime\prime}$ & 140000 & 2.3\textsuperscript{a} & \\
         $\mathrm{G173.4839}$& S233 IR\textsuperscript{f} & 05$^h$39$^{\prime}$09.92$^{\prime\prime}$ &+35$^{\circ}$45$^{\prime}$17.2$^{\prime\prime}$ & 2900 & 1.8\textsuperscript{b}\\
         $\mathrm{G192.6005}$ & S255\textsuperscript{g} &06$^h$12$^{\prime}$54.01$^{\prime\prime}$ &+17$^{\circ}$59$^{\prime}$23.1$^{\prime\prime}$ & 45000& 1.8\textsuperscript{c}\\
         $\mathrm{G196.4542}$& S269\textsuperscript{g} &06$^h$14$^{\prime}$37.06$^{\prime\prime}$ &+13$^\circ$49$^{\prime}$36.4$^{\prime\prime}$ & 94000 & 5.3\textsuperscript{d}\\
         \hline
         {\textsuperscript{a}\footnotesize{\citealt{2019MNRAS.487.2771N}}} & {\textsuperscript{b}\footnotesize{\citealt{1990ApJ...352..139S}}} & {\textsuperscript{c}\footnotesize{\citealt{2016MNRAS.460..283B}}} & {\textsuperscript{d}\footnotesize{\citealt{2007PASJ...59..889H}}} &
         {\textsuperscript{e}\footnotesize{\citealt{1972MNRAS.160....1W}}}\\
          {\textsuperscript{f}\footnotesize{\citealt{2016AstBu..71..208L}}} & {\textsuperscript{g}\footnotesize{\citealt{2002yCat.5112....0A}}}\\
         \multicolumn{4}{l}{\textsuperscript{*}\footnotesize{\href{http://rms.leeds.ac.uk/cgi-bin/public/RMS_DATABASE.cgi}{Red MSX Source Survey}}} 
    \end{tabular}
    }
    \label{Table_variability_sample}

\end{table*}

\subsection{VLA observation}
The four objects were observed on the 29 March 2018 using NRAO's JVLA telescope in A configuration at 6\,GHz. Two of them, G192.6005 and G196.4542, were observed again on 05 May 2018 at 22.2\,GHz, in the same configuration, to allow for estimation of spectral indices. The observations, done under the project code 18A-158, are of angular resolutions, $ 0.33^{\prime\prime}$  and $0.089^{\prime\prime}$;  bandwidths, 4 and 8\,GHz; and integration times, $\sim$ 20 minutes and 30\,minutes at 6\,GHz and  22.2\,GHz respectively. Besides the continuum emission, methanol and water masers of frequencies; 6.7\,GHz and 22.2\,GHz, bandwidths; 4.0\,MHz and 8.0\,MHz, and spectral channels; 256 and 128, respectively, were also observed to aid in the calibration of the data. A list of the observed sources and the calibrators used in scaling their fluxes and phases at 6.0 and 22.2\,GHz is shown in Table \ref{CnadKbandCalibrators}. The set of calibrators used in the 2012 observation were maintained to minimise comparison errors and maximise positional accuracy. 

\begin{table}
    \centering
        \caption{A table of sources and their calibrators. Fluxes for the phase calibrators, together with their angular sizes are shown on the table. $-$ means that a source is point-like. Fluxes of the sources were calibrated using 3C48.
        }
        
    \resizebox{\linewidth}{!}{%
    \begin{tabular}{c c c c c}
       \hline
         Source&  \multicolumn{4}{c}{Phase Calibrators}  \\
         
         \cline{2-5}
           & Phase Cal & $S_\nu$\,(mJy) & Size ($^{\prime\prime} \times ^{\prime\prime}$) & PA($^\circ$) \\
        \hline
        \multicolumn{2}{c}{} & At $\mathrm{6.0}$\,GHz  \\
        \cline{3-3}
        $\mathrm{G133.7150}$& J0228$+$6721 & 1280$\pm$5 & $-$ & $-$ \\
         $\mathrm{G173.4839}$  & J0555$+$3948 & 3980$\pm$10 &$0.05\pm0.01\times0.05\pm0.01 $& 170 $\pm$ 30\\
         $\mathrm{G192.6005}$  & J0559$+$2353 & 251$\pm$1 &$0.05\pm0.01\times0.04\pm0.01 $& 160 $\pm$ 10 \\
         $\mathrm{G196.4542}$  & J0559$+$2353 & 251$\pm$1 &$0.05\pm0.01\times0.04\pm0.01 $& 160 $\pm$ 10 \\
           &  &   &   &    \\

        \multicolumn{2}{c}{} & At $\mathrm{22.2}$\,GHz & \\
        \cline{3-3}

         $\mathrm{G192.6005}$ & J0559$+$2353 & 193$\pm$1 & $-$& $-$ \\
         $\mathrm{G196.4542}$ & J0559$+$2353 & 193$\pm$1 & $-$& $-$\\
         \hline
         
    \end{tabular}
    }
    \label{CnadKbandCalibrators}
\end{table}{}

Calibration and imaging were done using the NRAO's data reduction software,  CASA (Common Astronomy Software Applications; \citealt{2007ASPC..376..127M}). C-band images were created using Briggs \citep{1995AAS...18711202B} weighting of robustness 0.5, except in $\mathrm{G133.7150}$, where robustness was set at -1 due to the presence of diffuse emission of its field. In K-band, imaging of  $\mathrm{G192.6005}$ and $\mathrm{G196.4542}$ was done with robustness parameters of 0.0 and 0.5 respectively. $\mathrm{G192.6005}$ was in outburst and its field was affected by strong side-lobes which were suppressed by the Briggs weighting of robustness 0.0. The HPBW of the synthesised beams, used in deconvolving the dirty maps at both C- and K- bands, are shown in Table \ref{Table_synthesisedBeams}. Some of the images were generated after self-calibration of their visibilities, either once or twice, to improve their signal to noise ratios. Phase only self-calibration, and in some cases, followed by a combination of a phase and amplitude self-calibration were performed on the images. The resultant images have rms noise values shown in Table \ref{Table_synthesisedBeams}. 

The rms noise for three of the fields, at 6\,GHz, is 6\,$\mu$Jy/beam, however, in G133.7150's field, the rms is higher, 21\,$\mu$Jy/beam, due to extended emission in the field. At 22.2\,GHz, the rms noise is slightly higher than the rms noise at 6\,GHz i.e an average value of 10\,$\mu$Jy/beam. 

\begin{table}

\caption{Wavelength, synthesised beam, position angle and field rms of the observations.}
\resizebox{1.05\linewidth}{!}{%
\begin{tabular}{l l l l l l l l}
\hline
    Source & \multicolumn{6}{c}{Synthesised beams} \\
    \cline{2-7}
     & \multicolumn{3}{c}{$\mathrm{\nu=6\,GHz}$} &  \multicolumn{3}{c}{$\mathrm{\nu=22.2\,GHz}$} \\
    \cline{2-4}
    \cline{5-7}
    &  $\theta_{maj}\times \theta_{min}$ ($^{\prime\prime}$) & PA ($^\circ$) & rms($\mu$Jy/b)  & $\theta_{maj}\times \theta_{min}$ ($^{\prime\prime}$) & PA ($^\circ$) & rms($\mu$Jy/b)\\
\hline 

$\mathrm{G133.7150}$& $0.29 \times 0.19$ &  -64.1 & 21 & $-$ &  $-$ &  $-$ \\
$\mathrm{G173.4839}$& $0.29\times 0.26$ & -21.3  &   6 & $-$ &  $-$ &  $-$  \\
$\mathrm{G192.6005}$ & $0.29 \times 0.27$ &  15.1  &   6 &  $0.10 \times 0.09$& -51.23 & 9.5 \\
$\mathrm{G196.4542}$ & $0.29 \times 0.27$ &  4.3  &   6 &  $0.13 \times 0.10$& 76.45 & 11 \\

\hline
\end{tabular}
}
\label{Table_synthesisedBeams}
\end{table}

\section{Results}
\subsection{Continuum and Maser Emission}
All the sources were detected at the frequencies of the observations, five at 6\,GHz and two of them at 22.2\,GHz. At 6\,GHz, all the fields of the objects display evidence of multiple radio sources.  Eleven were detected in the field of G133.7150, three in both G173.4815 and G173.4839, nine in G192.6005 and four sources in G196.4542. In some cases, the 6\,GHz emission manifested elongated morphology, enclosing multiple components e.g G196.4542. G192.6005 and G196.4542, observed at both 6\,GHz and 22.2\,GHz, have radio sources that were detected at both frequencies. Four sources were detected at both frequencies within G192.6005's field and one in G196.4542's field. These sources show more compact  structures at 22.2\,GHz compared to 6\,GHz, even-though their emission at 22.2\,GHz also manifests extended features. In addition, both methanol and water maser species were detected in G192.6005 and G196.4542, however, the 6.7\,GHz methanol maser was absent in the fields of both G133.7150 and G173.4839 which were only observed at C-band.

Fluxes, sizes and positions of the sources were estimated using the CASA task $\tt{imfit}$.
Besides the uncertainties reported by $\tt{imfit}$, additional errors, added in quadrature, were included to account for errors due to calibration of the fluxes and positional accuracy of the observations. The uncertainties due to calibration of the fluxes were estimated to be 10\% of a source’s flux. Likewise, a typical value of the positional accuracy of the JVLA, estimated as 10\% of the HPBW of a synthesised beam, was used.
 
\subsection{Radio sources}
\subsubsection{G133.7150+01.215}
\begin{figure*}
    \centering
    \includegraphics[width = 0.75\textwidth]{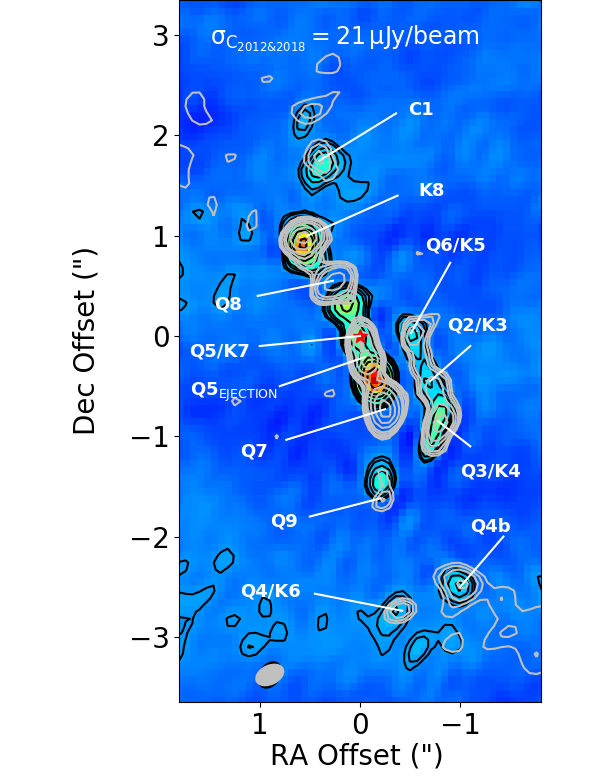}
    \caption{Colour map of G133.7150's field and its contours (shown in black) at C-band in the 2012 observation. The silver contours show the positions and morphologies of the radio sources in the 2018 observation. The contours levels are 3, 5, 7, 11, 15, 25, 37, 48 and 70$\sigma$ in both the 2012 and 2018 maps. The letters used in the naming of the radio sources correspond to the frequency band where they were first detected while the red asterisk represent the location of Q5/K7, the jet driver.}
    \label{G133_Cband}
\end{figure*}

G133.7150, also known as W3 IRS5, harbours compact radio sources, three of which have mid-IR counterparts \citep{2005A&A...431..993V}. All the radio sources detected in \citet{2005A&A...431..993V} and \color{blue}Purser et al. \color{black}(\color{blue}2020 in prep\color{black}) were detected at 6\,GHz except Q10 (\color{blue}Purser et al. 2020 in prep\color{black}). The labelling in \citet{2005A&A...431..993V} and \color{blue}Purser et al. \color{black}(\color{blue}2020 in prep\color{black}) was adopted here (see Figure \ref{G133_Cband}). Table \ref{Table_G133_sources} lists all the sources detected above 3$\sigma$. 

\begin{table}

\caption{Peak intensities, flux densities, deconvolved angular sizes and position angles of the sources detected within the field of G133.7150 at C-band in 2018. $-$ means that the source is a point source or its emission is elongated and irregular such that its actual size and orientation is not clear. The letter on the names indicates the frequency band at which the source was first detected.}
\resizebox{\linewidth}{!}{%
\begin{tabular}{c c c c c c c c}
\hline

    Source & Peak flux & S$\nu$ &  Major Axis &  Minor Axis & PA \\
     & (mJy/beam)  & (mJy)  & (arcsecs)  & (arcsecs) & (deg) \\
\hline 
C1&  0.18$\pm$0.03  &0.25$\pm$0.08 &  0.35$\pm$0.03 &  0.22$\pm$0.02 &  30$\pm$10 \\
Q2& 0.31$\pm$0.02 &  0.40$\pm$0.04 & $-$ &  $-$ &  $-$  \\
Q3& 0.65$\pm$0.03 &  0.80$\pm$0.05 & $-$ &  $-$ &  $-$  \\
Q6& 0.30$\pm$0.02 &  0.35$\pm$0.03 & $-$ &  $-$ &  $-$  \\

Q4&  0.25$\pm$0.03   &0.27$\pm$0.06& $-$ &  $-$ &  $-$ \\
Q4b& 0.24$\pm$0.04 &  0.59$\pm$0.07 & 0.36$\pm$0.09 & 0.17$\pm$0.09 &  41$\pm$25 \\
Q5 & 0.99$\pm$0.12  & 1.68$\pm$0.31 & $-$ & $-$ &  $-$ \\
Q7& 1.20$\pm$0.11  & 1.87$\pm$0.26 & $-$ & $-$ &  $-$ \\
K8 &0.98 $\pm$0.09 &  1.13$\pm$0.20& $-$ & $-$ &  $-$  \\
Q9 &0.16$\pm$0.02  & 0.21$\pm$0.10  &  $-$ & $-$ &  $-$  \\
Q8& 0.77$\pm$0.11 & 1.00$\pm$0.25 &    $-$ & $-$ &  $-$  \\

\hline
\end{tabular}
}
\label{Table_G133_sources}
\end{table}

\begin{figure*}
    \centering
    \includegraphics[width = 0.75\textwidth]{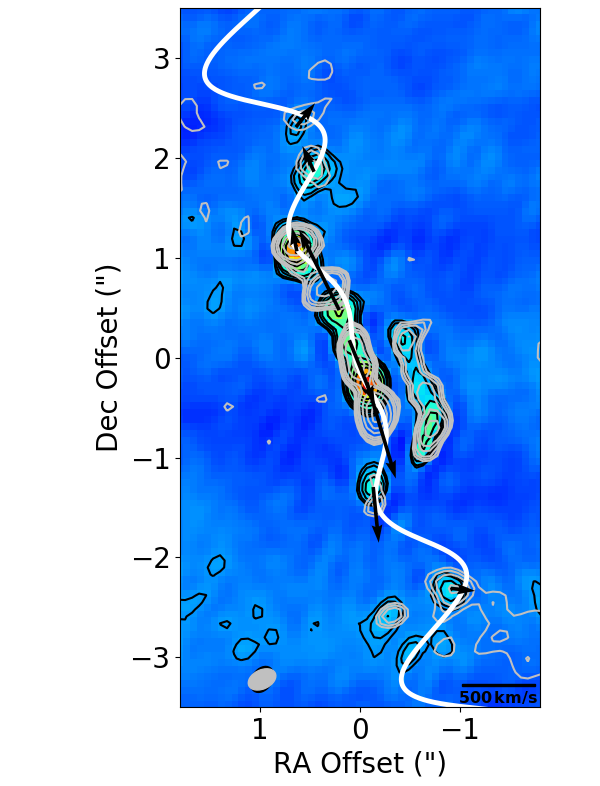}
    \caption{Precession model plotted over C-band contours of G133.7150 in white. Also included are the vectors of the proper motions which are shown as black arrows.}
    \label{proper_precess_lobesG133.png}
\end{figure*}

\citet{2005A&A...431..993V} established shifts in the positions of some of G133.7150's radio sources, translating to an average transverse velocity of $116$\,kms$^{-1}$, an estimate comparable to the proper motion of K8 \citep{2003ApJ...597..434W}. They also observed that the position angles (PAs) of the shifts are different, implying that G133.7150 drives a precessing jet.
Comparing the locations of some of the radio sources in 2012 (\color{blue}Purser et al. 2020 in prep\color{black}) with their positions in 2018, as measured here, confirms that they are moving away from Q5, perhaps the core of the jet (see Table \ref{Table_G133_sources_offset} and Figure \ref{proper_precess_lobesG133.png}). Indeed Q5 has a mid-IR counterpart and its position, at Q-band, in both \citet{2005A&A...431..993V} and \color{blue}Purser et al. \color{black}(\color{blue}2020 in prep\color{black}) only manifests a marginal offset of 0.04$^{\prime\prime} \pm$0.01 after a period of sixteen years. Moreover, C1, K8, Q7, Q8 and Q9 have negative spectral indices while Q5's spectral index is positive (\color{blue}Purser et al. 2020 in prep\color{black}), implying that Q5 is a thermal core, and C1, K8, Q7, Q8 and Q9, are its associated non-thermal lobes. Massive protostars with comparable morphologies were reported in \citet{2016MNRAS.460.1039P}, \citet{2019MNRAS.486.3664O} and \citet{2019A&A...623L...3S}. Q4b also has a negative spectral index and may be one of the lobes. Q4, on the other hand, has a mid-IR counterpart and its spectral index, 0.96$\pm$0.15, shows that it is thermal (\color{blue}Purser et al. 2020 in prep\color{black}), perhaps another core. Besides the sources in \citet{2005A&A...431..993V} and \color{blue}Purser et al. \color{black}(\color{blue}2020 in prep\color{black}), a new source, denoted as Q5$\mathrm{_{Ejection}}$ (see Figure \ref{G133_Cband}), was detected in the 2018 observation.  This source seems to be a new ejection from the core of the jet.

The information about the positional variability in G133.7150 can be used to estimate the proper motions of its lobes. The time gap of 1971 days permitted considerable angular displacements of the lobes, ranging from $\theta_s = $0.08$^{\prime\prime}$ to $\theta_s = $0.31$^{\prime\prime}$ as shown in Table \ref{Table_G133_sources_offset}. With the recent estimate of the distance to the object, $d=$ 2.3$\pm$0.2\,kpc \citep{2019MNRAS.487.2771N}, the lower limits of the proper motions of the lobes, calculated on the assumption that the jet axis is 90$^{\circ}$ to the line of sight, lie in the range; $v\sim$ 170$\pm$70\,kms$^{-1}$ for $\theta_s = $ 0.08$^{\prime\prime}\pm$0.01 and $\sim$ 650$\pm$50\,kms$^{-1}$ for $\theta_s = $ 0.31$^{\prime\prime} \pm$0.02. The average value of the proper motions of the lobes is $\sim$390$\pm$90\,kms$^{-1}$.

\begin{figure*}
    \centering
    \includegraphics[width = 0.75\textwidth]{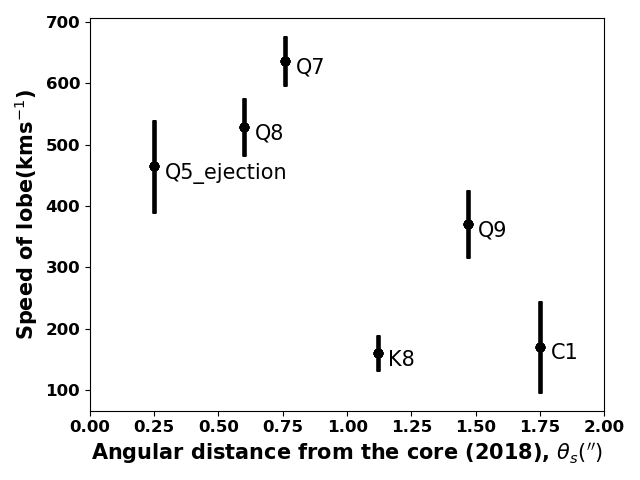}
    \caption{Proper motions of the different lobes of G133.7150+01.215.}
    \label{proper_Motions_lobesG133.png}
\end{figure*}

Figure \ref{proper_Motions_lobesG133.png} shows that the inner lobes have higher proper motions compared to the outer lobes. Moreover, their proper motions increase with distance, depicting an acceleration, especially near the core. This is also suggested by the non-detection of lobe d2 \citep{1994ApJ...424L..41C} in the 1996 observation \citep{2005A&A...431..993V}. The lobe, also denoted as Md2 in \citet{1997A&A...318..931T}, was detected by \citet{1994ApJ...424L..41C} and \citet{1997A&A...318..931T} at 15\,GHz and 5\,GHz respectively in 1989 but not in 1996 \citep{2005A&A...431..993V}, suggesting that it had moved to the position of K8 in $\lesssim$8 years, covering an angular distance of 0.99$^{\prime\prime}$ at an average velocity $v \geqslant 1300$\,kms$^{-1}$. The disappearance of d2 also implies that K8 is a terminal shock. Lobes K8, Q9 and C1, on the other hand, have lower velocities with no clear pattern, implying that they were decelerated, either due to a drop in their energies or as a result of interaction with ambient particles. Indeed, K8, Q9 and C1 seem most affected by the impact of the ambient medium, as they display the least displacement. For example, the proper motion of K8 is relatively stable. Its average velocity between 1989-1996 is 120$\pm$20\,kms$^{-1}$, a value that is comparable with its velocity between 2012-2018.

\begin{table*}

\caption{Displaced radio sources within G133.7150's field. Their locations, in 2012 and 2018, displacements, in arc seconds, and position angles of their proper motion are also shown.}
\resizebox{\linewidth}{!}{%
\begin{tabular}{c c c c c c c c}
\hline

    Source & \multicolumn{2}{c}{Position in 2012} &\multicolumn{2}{c}{Position in 2018} &  offset &  PA of PM  \\
\cline{2-5}
     & RA(J2000) & Dec(J2000) & RA(J2000) & Dec(J2000)  & ($^{\prime\prime}$) & (deg) \\
\hline 
C1& 02:25:40.835$\pm$0.002 & +062:05:54.276$\pm$0.016 & 02:25:40.840$\pm$0.003 &+62:05:54.353$\pm$0.025 &  0.08$\pm$0.03 & 24$\pm$22\\
Q5$_\mathrm{EJ}$ & 02:25:40.783$\pm$0.002& +062:05:52.542$\pm$0.007 & 02:25:40.771$\pm$0.002 & +062:05:52.328$\pm$0.031 &  0.23$\pm$0.04 & 200$\pm$7\\
K8 &02:25:40.861$\pm$0.001 & +062:05:53.452$\pm$0.007 & 02:25:40.861$\pm$0.002& +062:05:53.531$\pm$0.013 &  0.08$\pm$0.01 &5$\pm$12 \\
Q7& 02:25:40.760$\pm$0.001 & +062:05:52.139$\pm$0.006 & 02:25:40.746$\pm$0.001 & +062:05:51.840$\pm$0.017  &  0.31$\pm$0.02 & 198$\pm$5\\
Q8& 02:25:40.800$\pm$0.001 & +062:05:52.874$\pm$0.012 & 02:25:40.817$\pm$0.002 & +062:05:53.107$\pm$0.014   &  0.26$\pm$0.02 &27$\pm$5 \\
Q9 &02:25:40.749 $\pm$ 0.001 & +062:05:51.109$\pm$0.022 & 02:25:40.750$\pm$0.002 & +062:05:50.926$\pm$0.017 &  0.18$\pm$0.03  & 178$\pm$5\\

\hline
\end{tabular}
}
\label{Table_G133_sources_offset}
\end{table*}

The distribution of C1, Q4, Q7, Q8, Q9 and K8 on the sky creates a pattern that is point-symmetric about Q5. This, coupled with the directions of motion of C1, Q8, Q7 and Q9 suggests the presence of a precessing jet. A simple model, similar to the one described in \citet{1996AJ....112.2086E}, was adopted in assessing the nature of the precession. In this model, the path of a jet head  follows a conical spiral whose projection on the plane of the sky is a sinusoid. If the angle between the axis of precession and the line of sight is perpendicular, then the model can be described mathematically by equation \ref{jet_precess_model} \citep{1996AJ....112.2086E}, where $x$ and $y$ are the axial and radial displacements of the jet head from the location of the driving source respectively. $\phi$ is the position angle of the jet, $v$ is the velocity of the jet, $\alpha$ is the precession angle, $t$ is the time of evolution of the jet and $p$ is the period of precession.
\begin{equation}\label{jet_precess_model}
\begin{pmatrix}
x\\y
\end{pmatrix} = \quad
\begin{pmatrix}
\sin{\phi} & - \cos{\phi} \\
\cos{\phi} &  \sin{\phi}
\end{pmatrix}
\quad  
\begin{pmatrix}
vt \cos{\alpha}\\ vt \sin{\alpha} \sin{\left(\frac{2\pi}{p} t\right)}
\end{pmatrix}
\end{equation}

This simple model does not consider the angle of inclination of the jet to the plane of the sky and any variations in velocity along a jet, seen in G133.7150. Considering the wide range of the velocities of the lobes, only the first three lobes of the jet, perhaps the least affected by surroundings or loss of energy, was used in calculating the typical velocity of the jet's lobes, 540$\pm$50\,kms$^{-1}$. A constant velocity of 500\,kms$^{-1}$ and an inclination angle of 90$^{\circ}$ were therefore adopted in the calculations. The model, shown in Figure \ref{proper_precess_lobesG133.png}, was obtained for a jet whose position angle $\phi = 18^{\circ}\pm5$, precession period $p=40\pm3$\,years and precession angle $\alpha = 10^{\circ}\pm2$. This model, which we fit by inspection, seems to agree with the positions of the lobes and the directions of the displacement vectors despite its simplicity. The proper motion of some of the maser group B in \citet{2000ApJ...538..751I} also display comparable features. All the derived quantities are comparable to the calculations by \citet{2017PhDT.......147P} who used the position angles of the lobes to fit a precession model, however, \citet{2017PhDT.......147P}'s precession angle, 37$\pm$15, is higher than our estimate.

\begin{figure*}
    \centering
    \includegraphics[width = 0.72\textwidth]{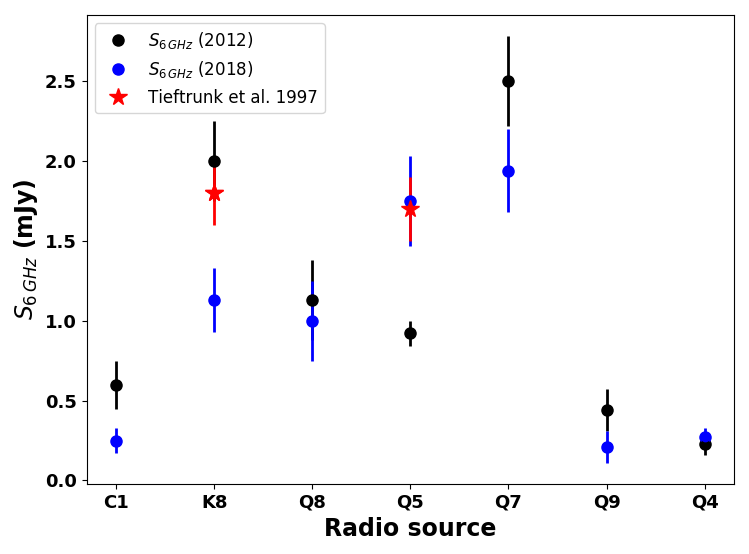}
    \caption{Flux densities of the radio sources associated with G133.7150. Blue data points represent the fluxes of the sources in 2018 while black points represent their peak fluxes in 2012 \citep{2017PhDT.......147P}.}
    \label{fig_fluxes_G133_20212_2018}
\end{figure*}

\citet{2005A&A...431..993V} noted a difference between the fluxes they derived and estimates from \citet{1997A&A...318..931T}, in some of the radio sources e.g Q5 and K8, suggesting flux variability. The flux densities of the radio sources in 2012 were compared with their fluxes in 2018 to identify the sources that experience flux variability. The differences, though marginal, is greatest in Q5 and K8 (see Figure \ref{fig_fluxes_G133_20212_2018}), perhaps due to changes during ejection of new lobes at Q5 and merging of lobes at K8. K8's flux was found to be lower in 2018 compared to its flux in both \citet{1997A&A...318..931T} and \color{blue}Purser et al. \color{black}(\color{blue}2020 in prep\color{black}). Q5's flux, on the other hand, is comparable to \citet{1997A&A...318..931T} in 2018 but lower than \color{blue}Purser et al. \color{black}(\color{blue}2020 in prep\color{black})'s estimate.


Radio sources Q2, Q3 and Q6 do not seem to be associated with the MYSO at Q5. Their alignment and morphology of emission depicts Q2 as a thermal jet, with Q3 and Q6 as its lobes. However, their spectral indices, 0.16$\pm$0.13, 0.76$\pm$0.10 and 0.58$\pm$0.20 (\color{blue}Purser et al. \color{black}\color{blue}2020 in prep\color{black}) respectively, suggest that they may be thermal cores. \color{blue}Purser et al. \color{black}(\color{blue}2020 in prep\color{black}) classified Q3 as a jet with lobes and both Q2 and Q6 as jet candidates. Indeed, the presence of water masers \citep{2000ApJ...538..751I} near the radio sources suggests that they are jets, however, only Q3 has a mid-IR counterparts and is a known core. 

\subsubsection{G173.4839+02.4317}
\begin{figure*}
    \centering
    \includegraphics[width = \textwidth]{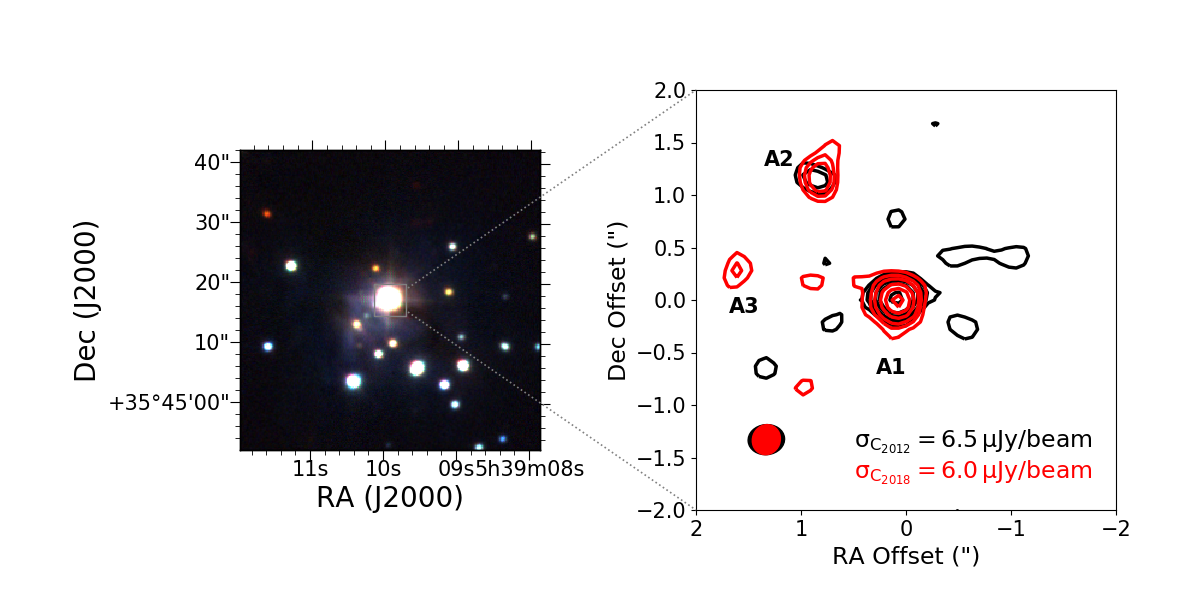}
    \caption{UKIDSS RGB colour (left) and C-band contour maps (right) of G173.4839. The black and red contours correspond to observations of 2012 \citep{2017PhDT.......147P}  and 2018 respectively. Contours levels are -3, 3, 5, 7, 9, 14, 17$\sigma$. The location of the (0,0) offset is ($\alpha=$ 05$^h$39$^{\prime}$09.92$^{\prime\prime}$, $\delta=$+35$^{\circ}$45$^{\prime}$17.2$^{\prime\prime}$).}
    \label{G173_4839Cband}
\end{figure*}

Two radio sources, A1 and A2, detected by \color{blue}Purser et al. \color{black}(\color{blue}2020 in prep\color{black}) in 2012, were detected in 2018 (see Figure \ref{G173_4839Cband}). A third source, A3, located to the east of A1, was also detected at 5$\sigma$. The flux densities of A1, A2 and A3 are 0.22$\pm$0.02\,mJy, 0.09$\pm$0.02\,mJy and 0.04$\pm$0.01\,mJy respectively. Neither A1 nor A2 manifest clear changes in both flux and positions. Identifying the nature of the sources is not straightforward from the limited information, however, their relative stability in both position and flux, low reddening seen in the 2MASS infrared image and bright Br$\gamma$ emission in A1 \citep{2013MNRAS.430.1125C}
suggests that they may be HII-regions (\color{blue}Purser et al. \color{black}\color{blue}2020 in prep\color{black}). 
The spectral indices (\color{blue}Purser et al. \color{black}\color{blue}2020 in prep\color{black}) of their emission also suggest that they are HII regions.
However, the 2.12\,$\mu$m H$_2$ map by \citet{2015MNRAS.450.4364N} detected a single outflow in the SE-NW direction. 

\subsubsection{G173.4815+02.4459}
\begin{figure*}
    \centering
    \includegraphics[width = \textwidth]{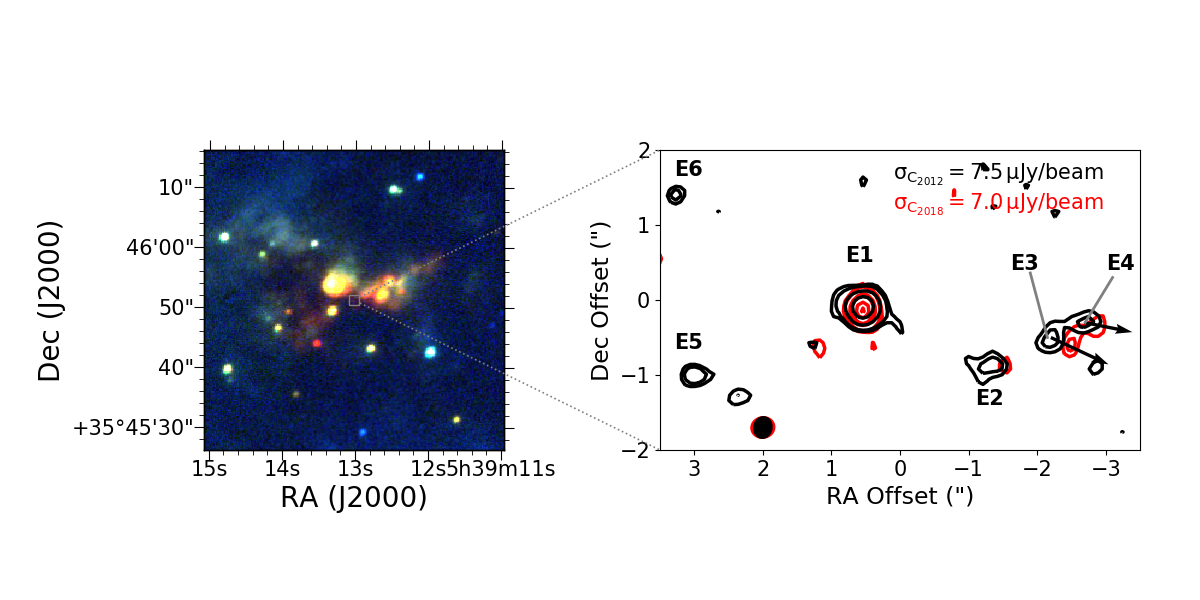}

    \caption{Colour map of UKIDSS RGB image of G173.4815. Contour maps of the fields at C-band in 2012 \citep{2017PhDT.......147P} and 2018 observations are also shown in black and red colours respectively. Contours levels are -3, 3, 5, 7, 9, 14, 17$\sigma$. The location of the (0,0) offset is ($\alpha=$ 05$^h$39$^{\prime}$13.02$^{\prime\prime}$, $\delta=$+35$^{\circ}$45$^{\prime}$51.3$^{\prime\prime}$).}
    \label{G173_4815_Cband}
\end{figure*}

The primary beam of G173.4839 at 6cm enabled the detection of another RMS MYSO, G173.4815, located at $\alpha = 05^h39^m13.02^{\prime\prime}	\, \delta = +35^d45^m51.3^{\prime\prime}$, approximately $52^{\prime\prime}$ away from the pointing centre of the observation. Three radio sources, E1, E3 and E4, shown in Figure \ref{G173_4815_Cband}, were detected in the field of the MYSO unlike in \color{blue}Purser et al. \color{black}(\color{blue}2020 in prep\color{black}) where six radio sources were detected. \color{blue}Purser et al. \color{black}(\color{blue}2020 in prep\color{black}) identified E1, a methanol maser source \citep{2000A&A...362.1093M}, as the core of an ionized jet while E3 and E4 were classified as lobes. Indeed, we measure the displacement of both E3 and E4 toward the west which implies that they are lobes and their driving sources are located to the east. This field is known to host three mm cores (\citealt{2002A&A...387..931B}, \citealt{2007A&A...466.1065B}) and multiple outflows, identified from CO and shock-excited H$_2$ emission (\citealt{2009ApJ...707..310G}, \citealt{2015MNRAS.450.4364N}, \citealt{2010MNRAS.404..661V}). One of the cores is co-located with E1 (\color{blue}Purser et al. \color{black}\color{blue}2020 in prep\color{black}), reaffirming that it is a core and the potential driver of the jet that harbours E3 and E4. The displacement vectors of E3 and E4, shown in Figure \ref{G173_4815_Cband}, suggest the presence of a precessing jet or multiple jet drivers. The magnitudes of their angular displacements are 0.30$^{\prime\prime}$ and 0.18$^{\prime\prime}$, translating to velocities of 530$\pm$50 and 320$\pm$30\,kms$^{-1}$ respectively, assuming that the distance of the source is 2.0$\pm$0.4\,kpc \citep{1998ApJS..117..387K} and that the motion is on the plane of the sky. 

Flux densities of the radio sources, derived from the 2018 observation, were found to be comparable to 2012 fluxes (\color{blue}Purser et al. \color{black}\color{blue}2020 in prep\color{black}) i.e 0.35$\pm$0.01\,mJy, 0.05$\pm$0.02 mJy and 0.04$\pm$0.02\,mJy for E1, E3 and E4 respectively.

\subsubsection{G192.6005-00.0479}

\begin{figure*}
    \centering
    \includegraphics[width = \textwidth]{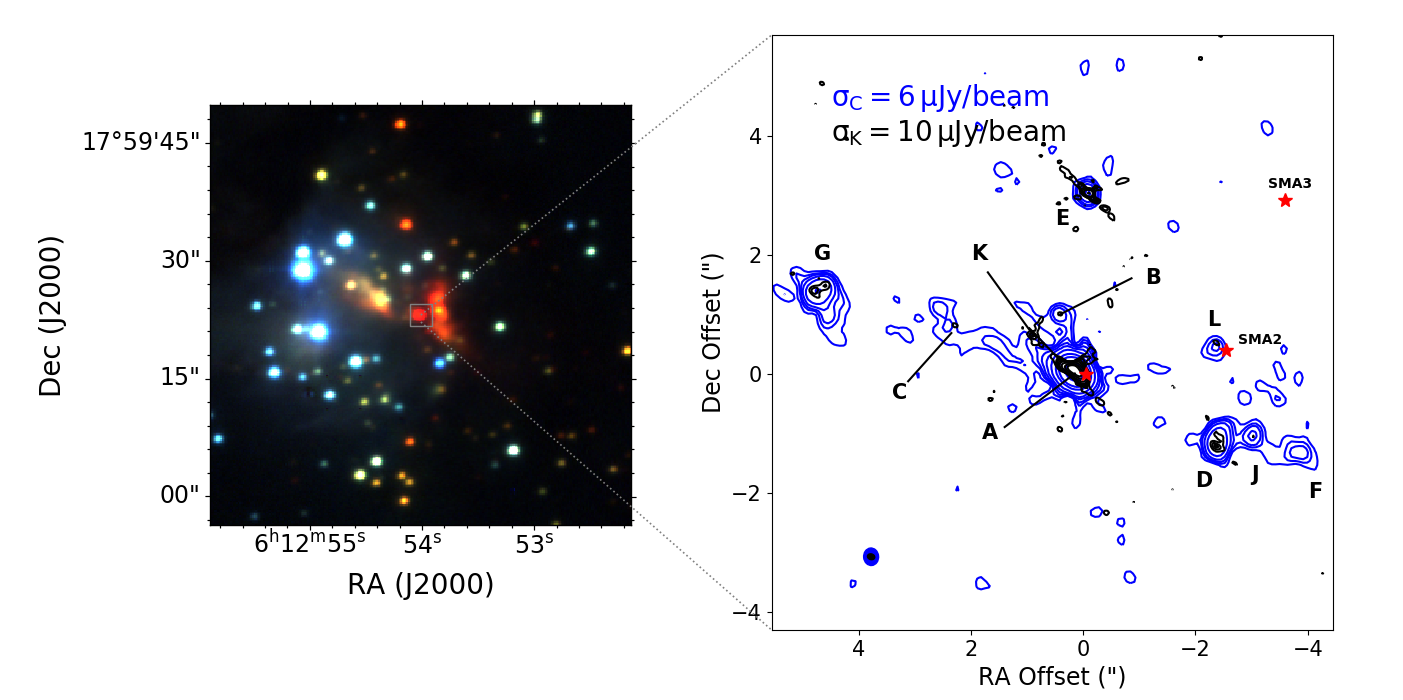}
    \caption{RGB colour image of G192.6005's field made from UKIDSS J,H,K images. Contours of C- and K-band emission of the field are also shown at levels 3, 5, 7, 11, 20, 50, 150, 300 \& 900$\sigma$, in blue and black colours respectively. The red asterisks represent locations of sub-millimetre cores in the field \citep{2018ApJ...863L..12L}. The location of the (0,0) offset is ($\alpha=$ 06$^h$12$^{\prime}$54.01$^{\prime\prime}$, $\delta=$+17$^{\circ}$59$^{\prime}$23.1$^{\prime\prime}$).}
    \label{G192_CandKbands}
\end{figure*}

G192.6005 is located in the  S255 star forming region. Its continuum map at both C- and K-bands exhibits the presence of a protostellar core, denoted as A (also known as S255 NIRS 3) and five jet lobes named C, D, F, G and J (see Table \ref{TableS255_at_6and1p3cm_Fluxes} and Figure \ref{G192_CandKbands}). The core of the protostars was resolved at both C- and K-bands, showing an elongation in a NE-SW direction at a position angle of 55$^\circ \pm$1. Its orientation is similar to the alignment of A, C, D, F, G and J. Moreover, the infrared emission in the field has a similar orientation, further confirming the presence of a jet in the NE-SW direction. Besides the protostar, there are other radio sources in the field, i.e B, E and K, that do not show any clear association with NIRS 3. B has a spectral index $\alpha>-0.58$, indicating that it is a non-thermal lobe while K's index $\alpha>1.33$ implying that it is a thermal source, perhaps an HII region. E, detected at both frequency bands has a spectral index of 1.18$\pm$0.08, another potential HII region. Only two of the radio sources in the field, A and L, have sub-millimetre counterparts \citep{2018ApJ...863L..12L} i.e thermal cores. L whose sub-millimetre counterpart is SMA2 was weakly detected at C-band and not K-band where rms noise is 10\,mJy/beam. A third sub-millimetre core, SMA3, does not have a radio counterpart at both C- and K-bands.

Previous C-band observation of the region by \citet{2017PhDT.......147P} revealed a similar distribution of the radio sources. The morphologies of the radio sources are also comparable. A and C, for example, are elongated in the NE-SW direction even-though C seems to be evolving into a new shape, perhaps due to an interaction between the fast moving particles of the jet and its slow moving clumpy counterparts. \citet{2018A&A...612A.103C} also detected the core, A, and its lobes but their images were more extended due to the lower resolution of their observations.\\

\begin{table*}
\caption{Positions, flux densities and peak intensities of the radio sources that were detected in the field of G192.6005 in 2018} 
\label{TableS255_at_6and1p3cm_Fluxes}

\begin{tabular}{ l l l c c c c c c c c c}
\hline
\multirow{2}{*}{Object} & \multicolumn{2}{c}{Position}& \multicolumn{2}{c}{Integrated Flux (mJy)} & \multicolumn{2}{c}{Peak flux (mJy/Beam)} & $\alpha_{_{CK}}$ & Nature\\
    \cline{2-3}
    \cline{4-5}
    \cline{6-7}
    \cline{8-9}

 & Dec & RA & C-band & K-band & C-band & K-band &  &  \\
    \hline
\cline{6-6}
A & 06$^h$12$^{\prime}$54.02$^{\prime\prime}$& +017$^\circ$59$^{\prime}$23.1$^{\prime\prime}$ & 15.11$\pm$0.09 & 24.72$\pm$0.80& $11.47\pm0.04$& $11.28\pm0.25$ & 0.37$\pm$0.02 & Core\\
B & 06$^h$12$^{\prime}$54.04$^{\prime\prime}$& +017$^\circ$59$^{\prime}$24.1$^{\prime\prime}$ & 0.13$\pm$0.03 & $<0.06$& $0.06\pm0.02$&$<0.12$ &$>-0.58$ & Lobe\\
D & 06$^h$12$^{\prime}$53.84$^{\prime\prime}$& +017$^\circ$59$^{\prime}$21.9$^{\prime\prime}$ & 0.55$\pm$0.03 & $0.27\pm0.04$ & 0.32$\pm$0.02& $0.12\pm0.03$ & $-0.53\pm0.12$ & Lobe \\
E & 06$^h$12$^{\prime}$54.00$^{\prime\prime}$& +017$^\circ$59$^{\prime}$26.1$^{\prime\prime}$ & 0.17$\pm$0.04 &$0.83\pm0.07$ & 0.15$\pm$0.02& $0.84\pm0.04$ & $1.18\pm0.08$ & HII-region\\
F & 06$^h$12$^{\prime}$53.74$^{\prime\prime}$& +017$^\circ$59$^{\prime}$21.78$^{\prime\prime}$ & 0.19$\pm$0.03 &$<0.08$ & 0.06$\pm$0.02&$ <0.12$ & $>-0.64$ & Lobe\\
G & 06$^h$12$^{\prime}$54.33$^{\prime\prime}$& +017$^\circ$59$^{\prime}$24.5$^{\prime\prime}$ & 0.65$\pm$0.04 &$0.26\pm0.05$ & 0.34$\pm$0.02& $0.11\pm0.03$ & $-0.68\pm0.15$ & Lobe\\
J & 06$^h$12$^{\prime}$53.80$^{\prime\prime}$& +017$^\circ$59$^{\prime}$22.1$^{\prime\prime}$ & 0.17$\pm$0.03 & $<0.07$& 0.07$\pm$0.02&$<0.12$ & $>-0.66$ & Lobe\\
K & 06$^h$12$^{\prime}$54.07$^{\prime\prime}$& +017$^\circ$59$^{\prime}$23.7$^{\prime\prime}$ & $<$0.05 &$0.30\pm0.06$ & $<0.12$ & $0.10\pm0.02$ & $>1.33$ & HII-region\\
L & 06$^h$12$^{\prime}$53.85$^{\prime\prime}$& +017$^\circ$59$^{\prime}$23.5$^{\prime\prime}$ & 0.09$\pm$0.02 &$<$0.07 & 0.05$\pm$0.01 & $<0.12$ & $<0.11$ & thermal core\\

\hline
\end{tabular}
\end{table*}

\noindent \textbf{G192.6005-00.0479's Core}

\begin{figure*}
    \centering
    \includegraphics[width = 0.495\textwidth]{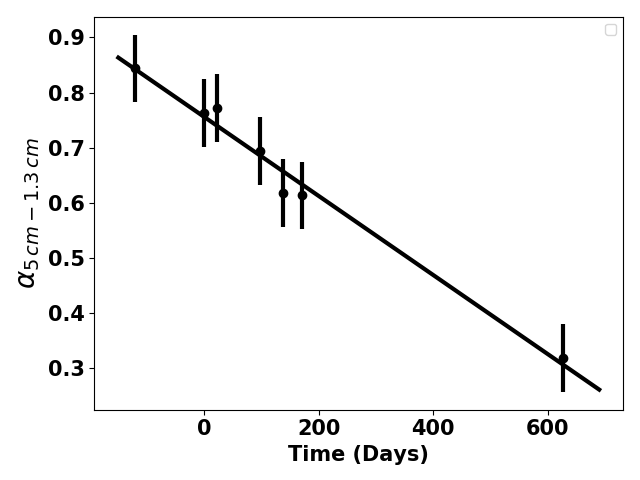}
    \includegraphics[width = 0.495\textwidth]{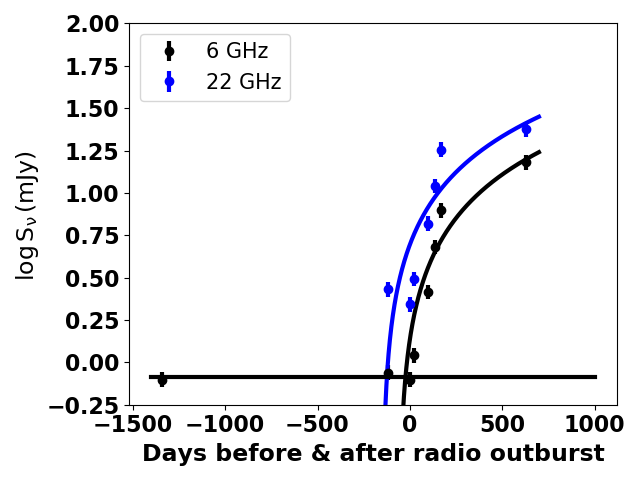}
    \caption{Evolution of the spectral index (left) and the fluxes (right) of S255 NIRS3 during outburst. Least square fits of the functions used to predict the trend of the fluxes and spectral index of the outburst are also shown in continuous lines. The horizontal line of the right panel is the approximate flux of the source at quiescence. Day zero corresponds to onset of the outburst.}
    \label{Nature_of_outburst}
\end{figure*}

\noindent The protostellar core of G192.6005, source A, also known as S255  NIRS 3, is a thermal radio source of spectral index 0.37$\pm$0.02. It displayed a significant rise in flux density after July 2016 \citep{2018A&A...612A.103C}. The flux density in 2012 was 0.79$\pm$0.02\,mJy \citep{2017PhDT.......147P}, a value comparable to \citet{2018A&A...612A.103C}'s estimates of March and July 2016. \citet{2018A&A...612A.103C} monitored the core from 11 March 2016 to 27 December 2016, a period totalling six epochs, following earlier reports of maser flaring \citep{2015ATel.8286....1F} and IR brightening \citep{2016ATel.8732....1S} of the source.  Their observations at both C- and K-bands manifest a clear increase in flux density of the core as shown by the right hand panel of figure \ref{Nature_of_outburst}. 

The data from \citet{2018A&A...612A.103C}, \citet{2017PhDT.......147P} and this observation, shown in Table \ref{Table_S255_fluxes}, were plotted to display the changes in the flux of the core for the period covering the eight epochs. As observed by \citet{2018A&A...612A.103C}, the core experienced an exponential rise in radio flux after July 2016. However, the observation of 2018 shows a clear deviation of the fluxes from the initial trend of the outburst. Whereas this deviation suggests that the outburst is either on a decline or plateauing, its methanol maser light curve \citep{2018A&A...617A..80S} points to a decline. More evidence of the decline was reported by \citet{2018ApJ...863L..12L} after analysing the source's 900\,$\mu$m continuum emission and
349.1\,GHz methanol maser emission.
\begin{table*}

\caption{Flux densities of the S255 NIRS3 at different epochs, from 2012 \citep{2017PhDT.......147P} to 2018. The 2016 data were taken from \citet{2018A&A...612A.103C}. The number of days before (negative) and after (positive) the radio outburst are also included assuming that 10 July 2016 was the onset of the outburst \citep{2018A&A...612A.103C}.}
\resizebox{\linewidth}{!}{%
\begin{tabular}{ l c c c c c c c c c c}
\hline
\multirow{2}{*}{Date} & \multicolumn{1}{c}{2012}& \multicolumn{6}{c}{2016} &   2018 \\
    \cline{2-2}
    \cline{3-7}
    \cline{8-9}
    \cline{10-10}
    
   &Nov. &  Mar.  & Jul. & Aug. & Oct & Nov. & Dec. &  Mar.  \\
 \hline
Days before \& after outburst   &-1342 & -121 & 0 & 22 & 97 & 137 & 170 & 627\\
\hline
Wavelength & (mJy)& (mJy)  & (mJy) & (mJy) & (mJy) & (mJy) & (mJy) & (mJy)  \\

\hline 
$\mathrm{\lambda = 1.34\,cm}$ & $-$  &2.7$\pm$0.3 &2.2$\pm$0.2 &3.1$\pm$0.3 &6.6$\pm$0.7 &11$\pm$1 &18$\pm$2 & 24.72$\pm$0.80\\
$\mathrm{\lambda = 5.00\,cm}$ &  0.79$\pm$0.08 &0.87$\pm$0.09 &0.79$\pm$0.08 &1.1$\pm$0.1 &2.6$\pm$0.3 &4.8$\pm$0.5 &7.9$\pm$0.8 & 15.11$\pm$0.09\\
\hline
\end{tabular}
}
\label{Table_S255_fluxes}
\end{table*}

Clearly, predicting the nature of the outburst is not straightforward given the large time-gap between the data from \citet{2018A&A...612A.103C} and the 2018 observation. To understand it further, the evolution of its spectral index was studied (see left panel of Figure \ref{Nature_of_outburst}). The spectral index of the central core, at each of the epochs, was derived using a combination of data from our 2018 observation and the C- and K-band data from \citet{2018A&A...612A.103C}. Uncertainties in the fluxes were all assumed to be 10\% of the flux for consistency with the errors used in \citet{2018A&A...612A.103C}. The index show a reduction with time: $\alpha_\nu=$0.84$\pm$0.06 in March 2016, 0.76$\pm$0.06 in July 2016 during the early stages of the outburst and 0.36$\pm$0.05 in 2018, over 600 days after the outburst.  Taking 10 July 2016 as the onset of the outburst i.e $t = 0$ \citep{2018A&A...612A.103C}, the spectral index $\alpha_\nu$ of the radio emission shows a linear relation with time $t$, $\alpha_\nu = -(6.5\pm0.5\times 10^{-4}) t + 0.75 \pm 0.01$. If the linear trend continues, the emission from the outburst should transition into an optically thin regime ($\alpha_\nu = -0.1$) after $\sim$ 1400 days unless a restoring mechanism acts. The continued downward trend in spectral index implies a drop in optical depth. Some of the factors that can result in the decline of the optical depth along a line of sight through the jet are: a drop in the jet number density $n$, a decrease in the path length through the jet $\Delta s$, a rise in the temperature $T$ of the jet and a drop in its ionization fraction $x$ \citep{1986ApJ...304..713R} i.e $\tau = \kappa_\nu \Delta s = a_\kappa n^2 x^2 T^{-1.35} \nu^{-2.1} \Delta s$, where $\kappa_\nu$ and $a_\kappa$ are the free-free absorption coefficient and 0.212 respectively. Effectively, the drop in optical depth suggests a number of possibilities which may include a combination of recombination of charged particles and/or a decrease in the jet density due to expansion and diffusion of jet particles. However, if the rise in flux is occasioned by an increase in the number density of the emitters, it would imply a higher opacity of the outburst. 
To explain the anti-correlation of the outburst's flux and spectral index, \citet{2018A&A...612A.103C} considered a model of an expanding gas bubble whose opening angle and the amount of ions injected into increases with time. While their model can explain the phenomenon, simulations of radio outbursts in galactic jets  (\citealt{2006A&A...456..895L}, \citealt{2000A&A...361..850T}) and the model of synchrotron and free-free emission radiated by a colliding wind binary \citep{2003A&A...409..217D} suggest that the S255 NIRS3 outburst could instead be due to a combination of synchrotron and free-free emission. The evolution of the synchroton emission, modelled in 3C 279 \citep{2006A&A...456..895L} and 3C 273 \citep{2000A&A...361..850T}, shows a rapid rise and then a decline. Such a rise in the intensity of synchrotron emission, if present in S255 NIRS3, could lower the overall index of the outburst. We therefore suggest that the anti-correlation of the intensity of S255 NIRS3's outburst with the spectral index of its emission may be an indirect indicator of the presence of non-thermal radiation from the core.

Evolution of the spectral index during the outburst shows a well-behaved linear correlation with time. This correlation can be used to predict the flux of the source at a given frequency using the equation $\log S_C = \log S_K +  \alpha_\nu(t) \log \left(\frac{\nu_{_C}}{\nu_{_K}}\right)$, where $S_C$, $S_K$, $\nu_{_C}$, $\nu_{_K}$, $\alpha_\nu(t)$ and $t$ are C-band flux, K-band flux, C-band frequency, K-band frequency, evolving spectral index and age of the outburst, respectively. 
Least-squares fitting of a logarithmic function of the form described above on the flux vs time data points displays a trend that points to a plateauing outburst.

\begin{figure*}
    \centering
    \includegraphics[width = 0.7\textwidth]{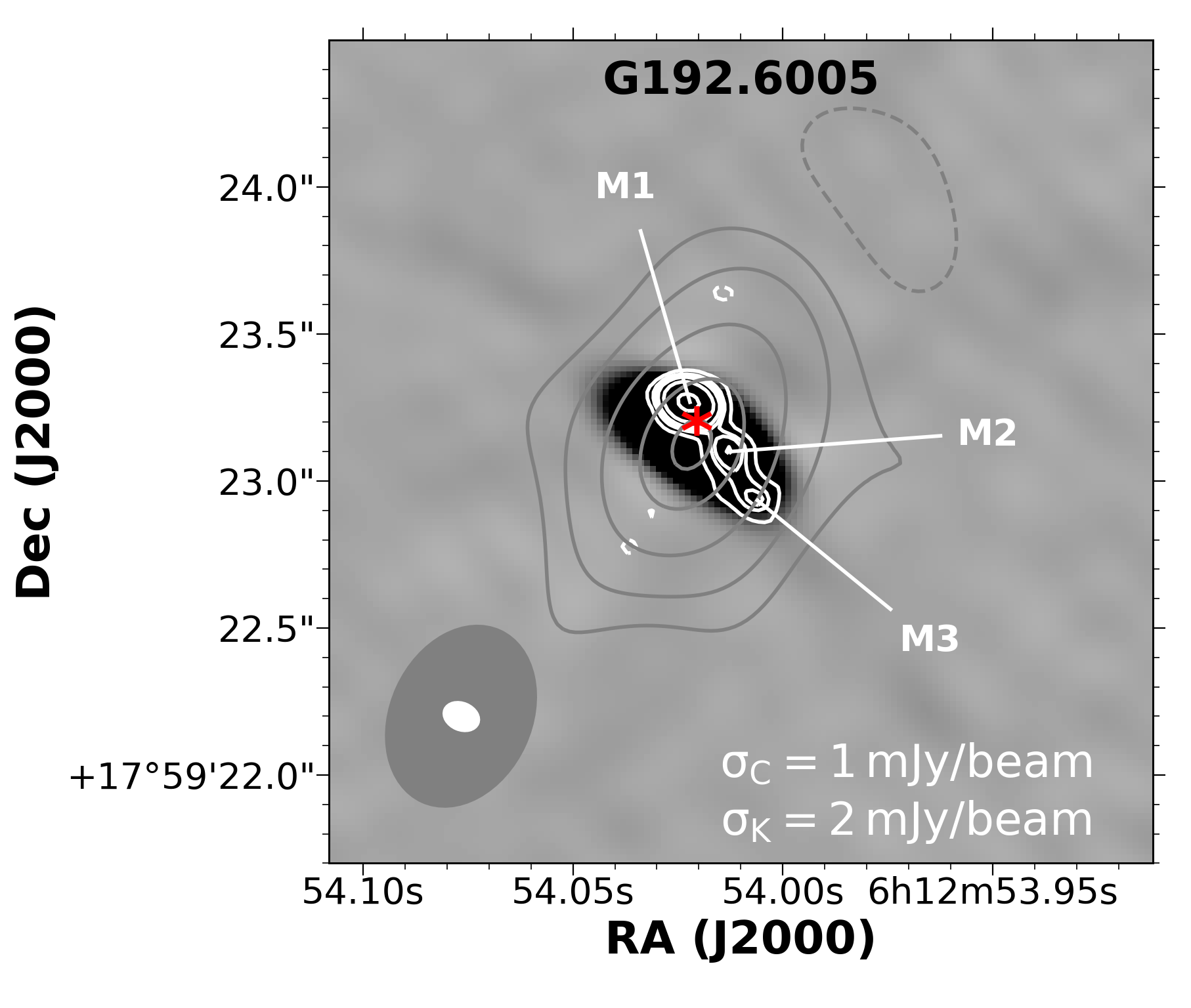}
    \caption{Greyscale map of G192.6005's continuum emission at 22.2\,GHz. Contour maps of 6.7\,GHz and 22.2\,GHz maser emission are also shown in gray and white colours respectively. The red asterisk represents the NIR K-band location of the core.}
    \label{Maser_K_band}
\end{figure*}

In addition to continuum emission, S255 NIRS 3 radiates 6.7\,GHz methanol maser emission and 22.2\,GHz water maser emission. Contour maps of the intensities of the water and methanol maser emission for the 2018 observation are shown in Figure \ref{Maser_K_band}. The methanol maser is marginally resolved  and does not show a clear morphology, however, the water masers trace a resolved structure that is well-aligned with the continuum emission. \citet{2017A&A...600L...8M} compared the locations of the 6.7\,GHz methanol maser, before and during outburst, and noticed that most of the maser locations were farther away from the core during outburst, implying a change in the circumstellar environment of the source \citep{2018A&A...612A.103C}. \citet{2018A&A...617A..80S} analysed data, collected through long-term monitoring of 6.7\,GHz methanol maser emission from S255 NIRS 3, reporting a rise in its intensity in two epochs. 
In the 2018 observation, the flux density of the emission is 43.4$\pm$2.7\,mJy, a value lower than the average flux of the radiation during quiescence. The intensity map of the water masers displays three peaks which are potential locations of the jets' knots. The peaks are arranged in a linear fashion with angular separation between M1, M2 and M3 displaying a quasi-regular pattern, implying that the processes that are responsible for their formation may be semi-regular; i.e., periodic ejection or precession. 

The mass-loss rate of S255 NIRS3 was calculated using the C-band flux of 2018 and compared with its rate before the outburst (\color{blue}Purser et al. \color{black} \color{blue}2020 in prep\color{black}). In the calculations, the jet parameters that were adopted in \citet{2018A&A...612A.103C} were used i.e a terminal velocity $v_t =900$\,km\,s$^{-1}$, ionisation fraction $x_o = 0.14$ and an inclination angle of 85$^o$, giving a comparable mass loss rate of $\dot M = 3.1\pm0.2 \times 10^{-5}$\,$\mathrm{M_\odot yr^{-1}}$. This rate is approximately an order of magnitude higher than its rate during quiescence i.e $2.91\pm2.19\times 10^{-6}\,\mathrm{M_\odot yr^{-1}}$ (\color{blue}Purser et al. \color{black} \color{blue}2020 in prep\color{black}). Assuming $\frac{M_{out}}{M_{acc}} \sim 0.1$ implies an accretion rate $\dot M_{acc} = 3.1 \times 10^{-4}$\,$\mathrm{M_\odot yr^{-1}}$ which is lower than the estimate from \citet{2017NatPh..13..276C}, suggesting that the factor of 10\%  under-estimates accretion rates. Indeed \citet{1994ApJ...436..125H} finds the ratio to be $\sim$1\% in low mass stars, a value that generates a rate that is comparable to \citet{2017NatPh..13..276C}.\\

\noindent \textbf{Radio lobes in G192.6005-00.0479}

\noindent Four radio sources, D, F, G and J, were found to have negative spectral indices, typical of non-thermal lobes. They are also aligned in a direction similar to the orientation of the radio emission of A and infrared emission in the field. This implies that they are lobes of a jet emanating from core A. Sources B and K, located to the north of A, are its closest neighbours, however, the location of B and spectral index of K are inconsistent with lobes of a jet driven by A. B is misaligned with the jet and K is a thermal radio source of spectral index $\alpha > 1.33$. The spectral index of B, $\alpha >-0.58$, shows that it is a non-thermal lobe, however, its associated core is unclear. 

\begin{figure*}
    \centering
    \includegraphics[width = 0.495\textwidth]{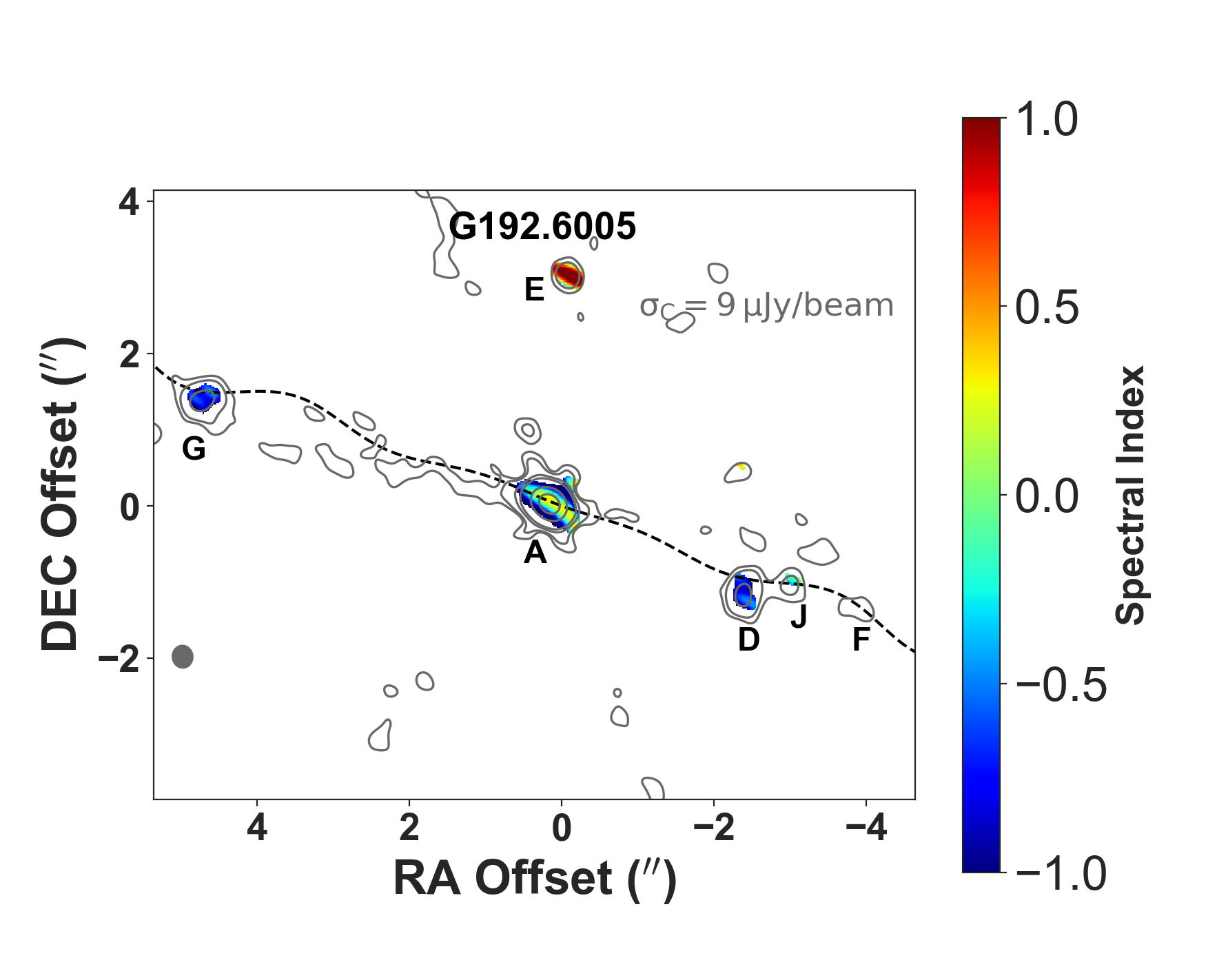}
    \includegraphics[width = 0.495\textwidth]{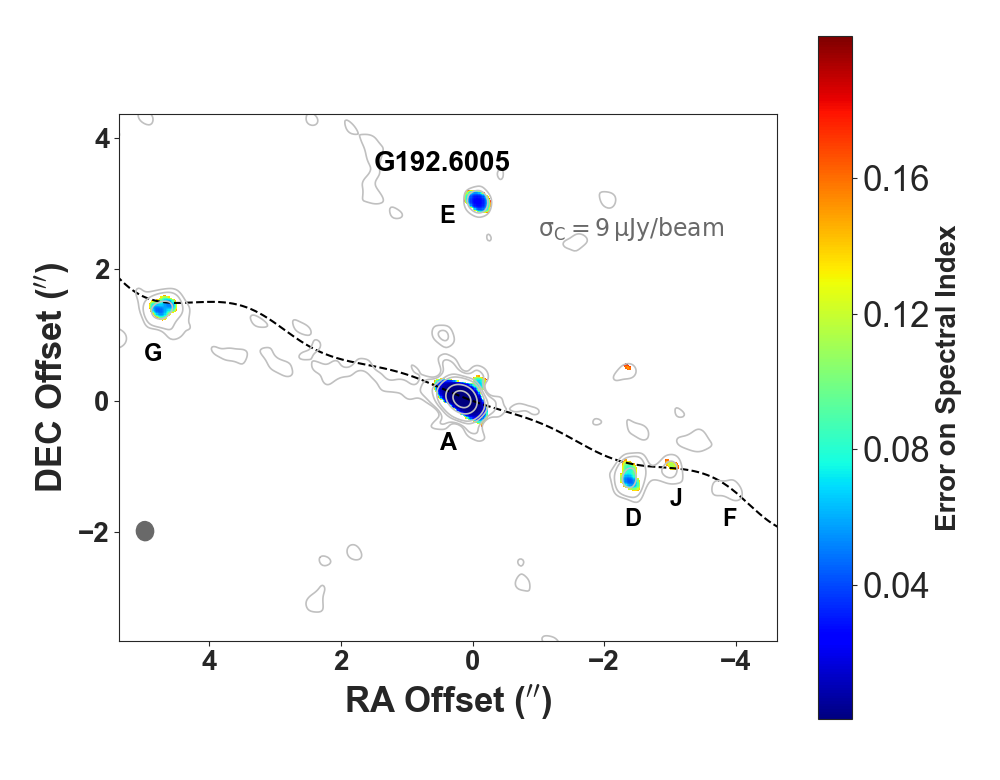}
    \caption{Spectral index map (left) and corresponding errors (right) of the field. The indices were only calculated if the flux of a pixel is higher than 3$\sigma$ in both frequency bands. The dashed lines represent a precession model in which the jet is perpendicular to the line of sight.}
    \label{G192_spix_map}
\end{figure*}

A spectral index map of the field was generated from C- and K- band data of similar uv coverage. The full uv-coverage of C- and K- band data are $11-950$\,k$\lambda$ and $37-2600$\,k$\lambda$ respectively, however, the overlap which occurs at 38-950k$\lambda$ was used. The map, shown in Figure \ref{G192_spix_map}, reveals the thermal core A and two non-thermal lobes, G and D. The orientations of G and D, at K-band, are perpendicular to the direction of the flow of jet particles, $\overrightarrow{AD}$ and $\overrightarrow{AG}$. 
The misalignment of some of the radio lobes, e.g, D, J and F, also suggests precession. The precession model in equation \ref{jet_precess_model} for a jet of precession angle $\alpha =2^{\circ}$ and period $p=$50 years seems to fit the distribution of the radio sources (see Figure \ref{G192_spix_map}), assuming that the jet drives out mass at a velocity $v=$500\,kms$^{-1}$ and has an inclination angle 90$^{\circ}$.

Considering that the bow shocks, G and D, may be internal and not terminating shocks, it is not possible to estimate the kinematic length and age of the jet with certainty. However, the angular separations of D and G from core A are $\sim$ 2.5$^{\prime\prime}$ and 4.8$^{\prime\prime}$ respectively, translating to physical distances of 0.02\,pc and 0.04\,pc respectively. Assuming a jet velocity of 500\,kms$^{-1}$ implies that it took 39 years and 78 years for the jet particles to travel from the core to D and G respectively. Their minimum energy and minimum magnetic field strengths, calculated from the equipartition theorem \citep{2014ApJ...792L..18A} are 4.5$\times$10$^{41}$\,ergs and 1.1\,mG respectively. In both F and J, the minimum B-field is 0.8\,mG while their minimum energies are 2.3$\times$10$^{41}$\,ergs and 2.1$\times$10$^{41}$\,ergs respectively.

\begin{figure*}
    \centering
    \includegraphics[width = 0.48\textwidth]{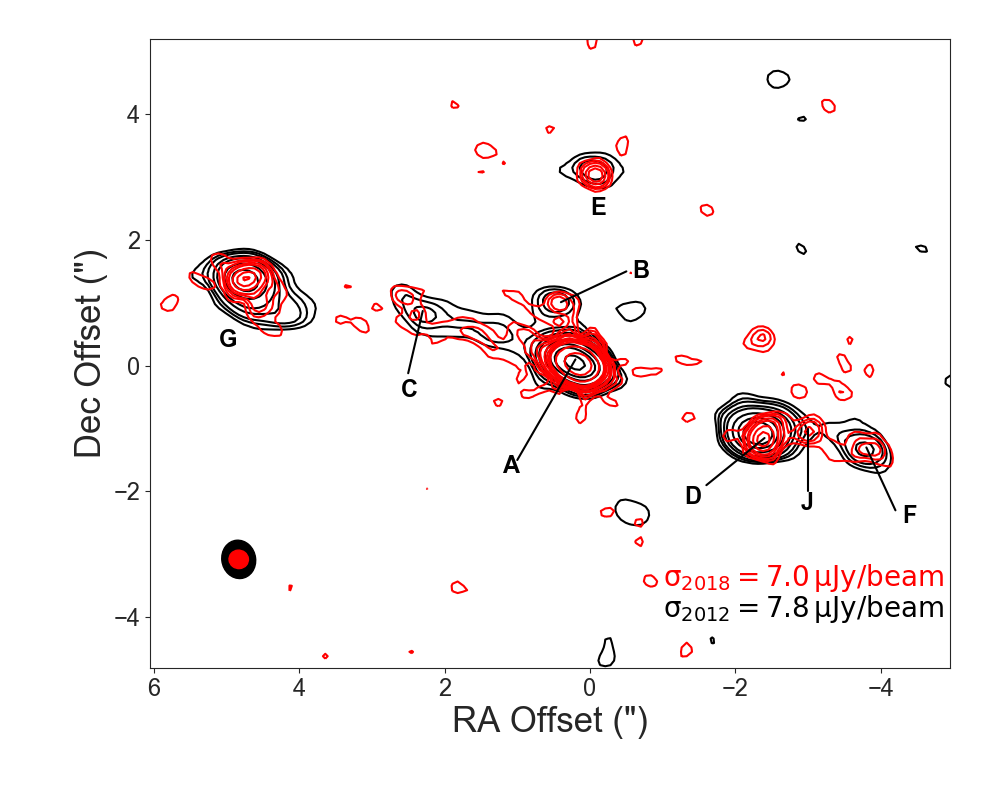}
    \includegraphics[width = 0.51\textwidth]{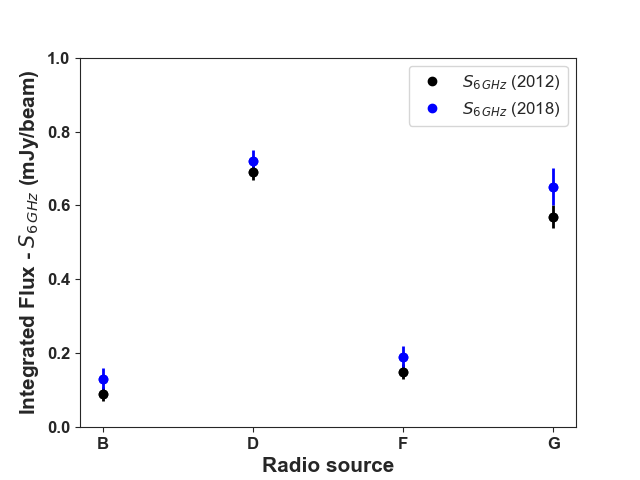}
    \caption{Left: Comparing the locations of the radio sources in G192.6005-00.0479. Their positions in 2012 are shown in black contours while 2018 locations are represented by the red contours. Right: Comparing the C-band fluxes of the radio lobes that are associated with core A. Blue dots represent the fluxes in 2018 while the black dots represent 2012 fluxes.}
    \label{G192_positional_variability.png}
\end{figure*}

The peak emissions of C and D in 2018 displayed slight angular displacements from their 2012  positions, however, the fluxes of all the lobes did not show any significant change (see Figure \ref{G192_positional_variability.png}). Adjacent to lobe D is a new lobe labelled J. This lobe was not detected in 2012 \citep{2017PhDT.......147P} and may be a new condensation, stemming out of D. The flux of D in 2012 was compared with the sum of the fluxes of D and J in 2018 since J may be part of D in the 2012 observation. The angular displacements of the peaks of C and D were found to be 0.2$^{\prime\prime}$ and 0.14$^{\prime\prime}$, translating to proper motions of 316$\pm30$\,kms$^{-1}$ and 220$\pm$20\,kms$^{-1}$ respectively, at the distance of the core, 1.8\,kpc. 

\subsubsection{G196.4542$-$01.6777}
\begin{figure*}
    \centering
    \includegraphics[width = \textwidth]{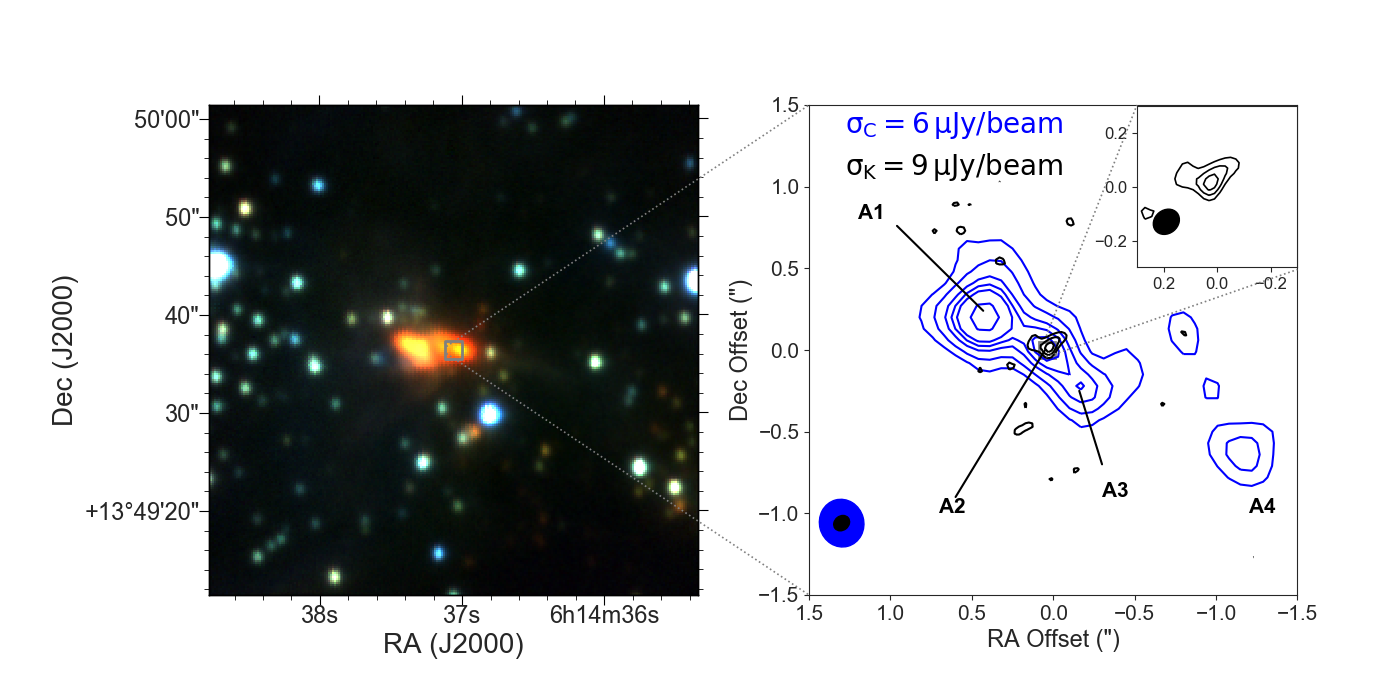}
    \caption{Left: RGB map of G196.4542's UKIDSS J, H, K images. Right: C- and K-band contours are of levels; -3, 3, 7, 11, 14, 17, 25$\sigma$ and -3, 3, 6, 9$\sigma$, shown in blue and black colours respectively. The location of the (0,0) offset is ($\alpha=$ 06$^h$14$^{\prime}$37.06$^{\prime\prime}$, $\delta=$+13$^{\circ}$49$^{\prime}$36.4$^{\prime\prime}$).}
    \label{G196_IR_CandKbands}
\end{figure*}

\begin{figure*}
    \centering
    \includegraphics[width = 0.8\textwidth]{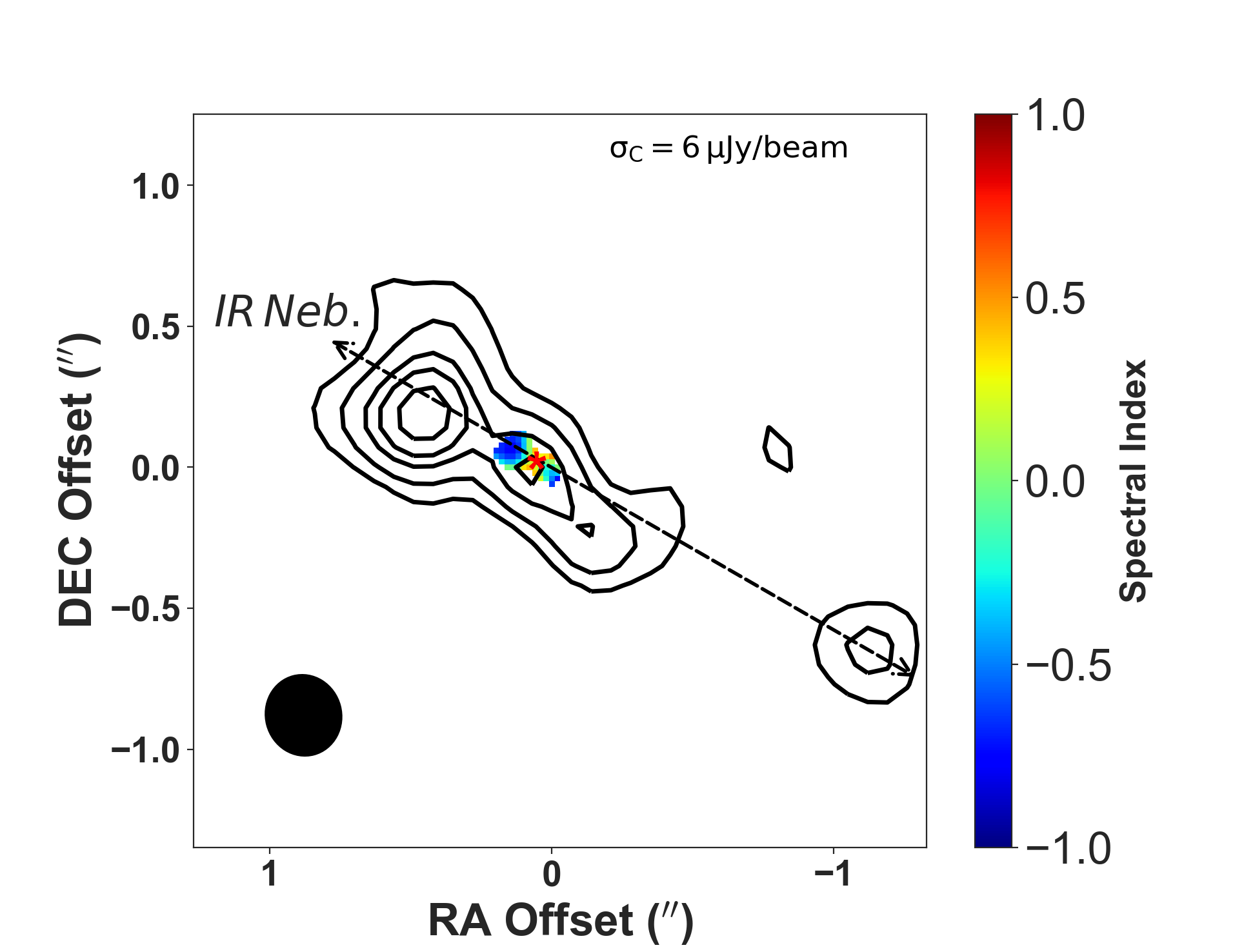}
    \includegraphics[width = 0.8\textwidth]{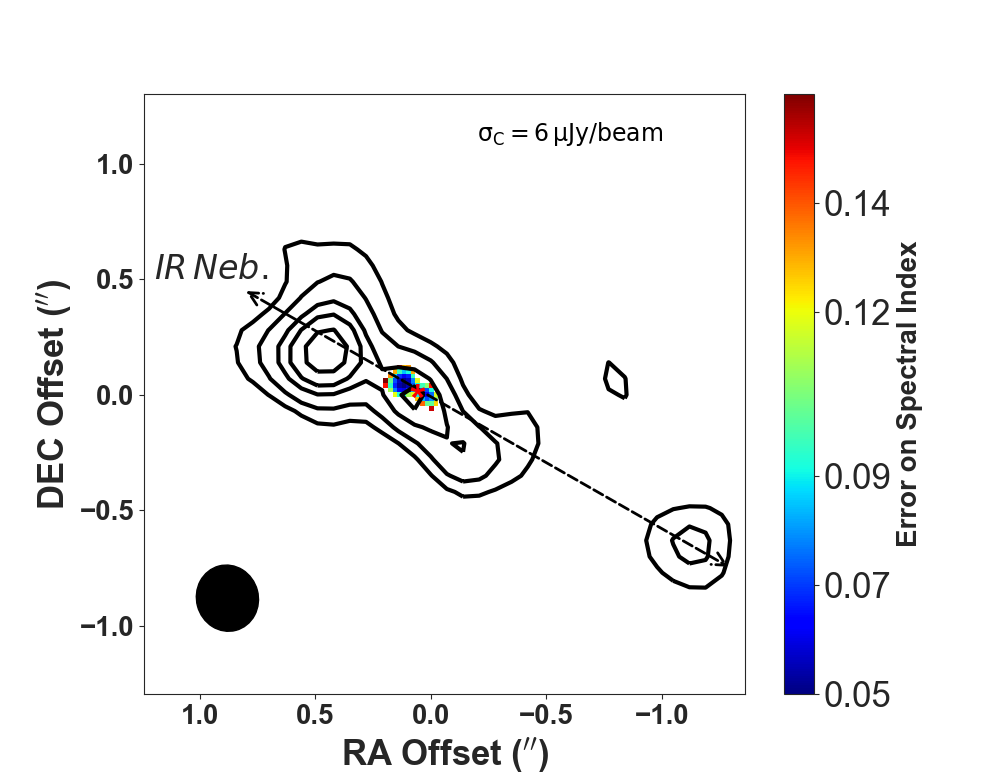}
w    \caption{Spectral index (top) and error (bottom) maps of G196.4542's field. C-band contours of levels; -3, 3, 7, 13, 18, 25, 50$\sigma$ are also shown. The red asterisk represents the NIR K-band location of the core.}
    \label{G196_SpixPlot_anderror}
\end{figure*}

Our C-band map of this source, the star-forming region S269 \citep{2003ApJ...596.1064J}, is shown in  Figure \ref{G196_IR_CandKbands}. It shows emission from four compact radio sources, A1-A4, with a weak extended emission surrounding A1-A3. 
 The orientation of the C-band emission is comparable to the alignment of infrared emission in the field, suggesting that A1, A2, A3 and A4 are components of a jet aligned in the same direction. A near infrared study of the region by \citet{2003ApJ...596.1064J} also detected H$_2$ emission knots that are aligned in a NE-SW direction. A2 was detected at both C- and K-bands while A1, A3 and A4 were only detected at C-band, implying that A2 is the core of the jet and A1, A3 and A4 are its non-thermal lobes.  The spectral index map of the field, derived from images of similar uv-coverage, also confirms that A2 is the core while the rest are non-thermal lobes (see Figure \ref{G196_SpixPlot_anderror}). A list of the radio sources, their flux densities and spectral indices is given in Table \ref{G196_Table_of_fluxes_and_spix}.  The radio sources do not show significant changes in flux or position except the flux of A1 which dropped by $\sim$25\%.

\begin{figure*}
    \centering
    \includegraphics[width = 0.8\textwidth]{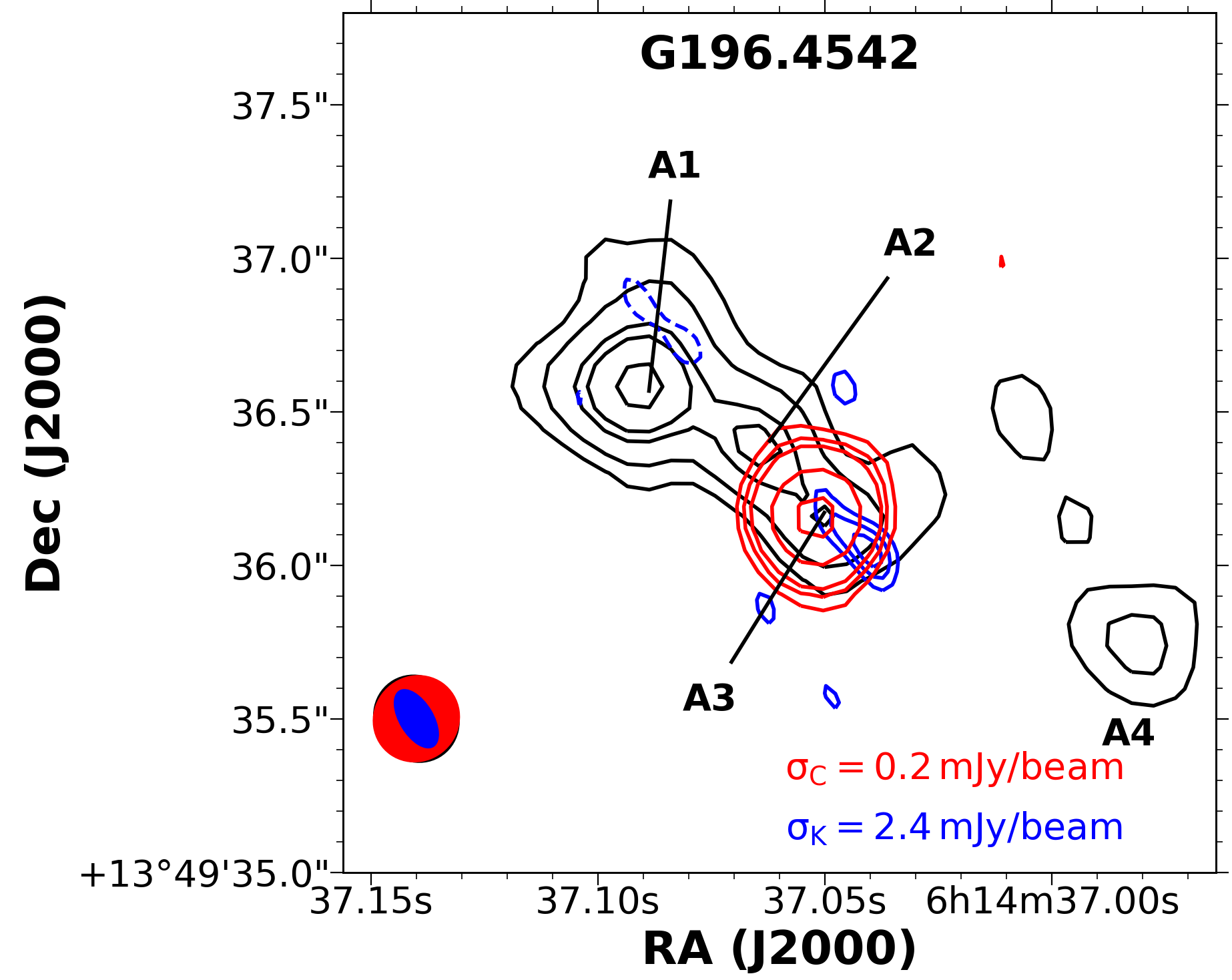}
    \caption{Contour maps of C-band continuum (black), 6.7\,GHz methanol masers (red), and 22.2\,GHz water masers (blue) emission of G196.4542. Their contour levels are -3, 3,  7, 14, 18 \& 27$\sigma$; -3, 3, 5, 7, 15 \& 25$\sigma$, and -3, 3, 5, \& 7$\sigma$ respectively }
    \label{Fig_G196_maser_emission}
\end{figure*}

Besides continuum emission, radiation from methanol (6.7\,GHz) and water (22.2\,GHz) masers was also detected. The methanol and water maser's flux densities are 6.4$\pm$0.4\,mJy and 34$\pm$7\,mJy respectively. Their peaks are located $\sim$0.3\,$^{\prime\prime}$ and 0.5\,$^{\prime\prime}$ to the SW of the core. The peak of the 6.7\,GHz emission coincides with the location of A3 while that of the 22.2\,GHz emission manifests an elongation that extends from A3 towards the SW (see Figure \ref{Fig_G196_maser_emission}). These masers appear to trace the the outer parts of the jet, away from the core.

\begin{table*}
\resizebox{\linewidth}{!}{%
\begin{threeparttable}
\caption{C-band positions, flux densities and spectral indices of radio sources within the field of G196.4542} 
\label{G196_Table_of_fluxes_and_spix}

\begin{tabular}{ l l l c c c c c c c c c}
\hline
\multirow{2}{*}{Object} & \multicolumn{2}{c}{Position}& \multicolumn{2}{c}{Integrated Flux (mJy)} & $\alpha_{_{CK}}$ & Nature\\
    \cline{2-3}
    \cline{4-5}
    \cline{6-7}
    \cline{8-9}

 & Dec & RA & C-band & K-band  &  &  \\
    \hline
\cline{6-6}
A1& 06$^h$14$^{\prime}$37.09$^{\prime\prime}$& +013$^\circ$49$^{\prime}$36.6$^{\prime\prime}$ & 0.30$\pm$0.02 & $<$0.08 &  $<$-1.33 & Lobe \\
A2 & 06$^h$14$^{\prime}$37.06$^{\prime\prime}$& +013$^\circ$49$^{\prime}$36.4$^{\prime\prime}$ & 0.10$\pm$0.02 & 0.13$\pm$0.02 & $0.20\pm0.19$ & Core \\
A3 & 06$^h$14$^{\prime}$37.05$^{\prime\prime}$& +013$^\circ$49$^{\prime}$36.2$^{\prime\prime}$ & 0.09$\pm$0.02 & $<0.05$ &  $<-0.44$ & Lobe \\
A4 & 06$^h$14$^{\prime}$36.98$^{\prime\prime}$& +013$^\circ$49$^{\prime}$35.8$^{\prime\prime}$ & 0.07$\pm$0.02 &$<0.05$ &  $<-0.25$ & Lobe \\

\hline
\end{tabular}
\end{threeparttable}
}
\end{table*}

\section{Conclusions}
We present new JVLA observations of five protostellar jets at C- and K-bands. Four of the jets manifest variability in flux, position or both except G173.4839. All the jets have thermal cores  whose positions are relatively fixed. However, their lobes are transitional objects that undergo noticeable positional variability. The proper motions of the jet knots/lobes show a wide range of velocities. The minimum and maximum estimates of their proper motions were found to be 170$\pm$70\,kms$^{-1}$ and 650$\pm$50\,kms$^{-1}$ respectively. In some jets, e.g. in $G133.7150$, the proper motion of the lobes was found to show a spread, implying that the lobes undergo periods of acceleration and deceleration as they move away from the core. A keen look at the change in the position of the lobes suggests that they move both radially and tangentially manifesting precession, e.g in $G133.7150$ and $G192.6005$, most likely due to close binary interaction.

The fluxes of both cores and lobes of some of the jets also displayed variability. Flux variability in the cores can be attributed to accretion activities while in lobe, diffusion and recombination of ionized particles can result in lower fluxes. Lobe K8 of  G133.7150 displayed the largest change in flux, perhaps due to an instance of merger with a subsequent lobe. Conversely, higher magnetic flux density and energised flow can result in a higher radio flux. Of the four MYSOs, flux variability was strongest in S255 NIRS3 which was in outburst. The opacity of its emission was found to decrease with time, consistent with a model of an expanding jet bubble. Similarly, emission from a new non-thermal lobe that is emerging from a core whose emission enshrouds that of the lobe has the potential to account for the increase in flux and a decrease in the spectral index of the outburst.
Predictions suggest that the outburst is either plateauing or declining. The mass-loss rate  \citep{1986ApJ...304..713R} of S255 NIRS3 during outburst shows that massive protostars are capable of attaining the high accretion rates \citep{1994ApJ...436..125H} needed to overcome radiation pressure through accretion bursts.

\section{Acknowledgement}
Special thanks to the UK's Science and Technology Facilities Council (STFC) who supported the research through WOO's PhD studentship under the DARA project-grant number ST/M007693/1. We are also grateful to Prof R. Cesaroni for sharing their January 2018 observation with us. 

\section{Data Availability}
Data Availability Statement: The data underlying this paper are available at the JVLA Data Archive: \url{https://archive.nrao.edu/archive/advquery.jsp}
\bibliographystyle{apj}
\bibliography{references}

\label{lastpage}
\end{document}